\documentclass[preprint,12pt]{elsarticle}

\usepackage{amssymb}

\usepackage{epsf}
\usepackage{graphicx} 
\usepackage{color} 
\usepackage{ulem}
\usepackage{amsfonts}
\usepackage{amsmath,amssymb} 
\usepackage{subfigure}
\usepackage{color}
\usepackage{multirow}
\biboptions{sort&compress}

\journal{Journal of magnetism and magnetic materials}

\begin{document}

\begin{frontmatter}



\title{The Spin-Half {\it XXZ} Antiferromagnet on the Square Lattice Revisited: A High-Order
Coupled Cluster Treatment}



\author[manchester]{R.F. Bishop}
\ead{raymond.bishop@manchester.ac.uk}
\author[manchester]{P.H.Y. Li}
\ead{peggyhyli@gmail.com}
\author[magdeburg1]{R. Zinke}
\ead{RonaldQZinke@web.de}
\author[magdeburg1]{R. Darradi}
\author[magdeburg1]{J. Richter}
\ead{Johannes.Richter@physik.uni-magdeburg.de}
\author[cardiff]{D.J.J. Farnell}
\ead{FarnellD@cardiff.ac.uk}
\author[magdeburg2]{J. Schulenburg}
\ead{Joerg.Schulenburg@URZ.Uni-Magdeburg.DE}

\address[manchester]{School of Physics and Astronomy, Schuster Building, The University of Manchester, Manchester, M13 9PL, UK}
\address[magdeburg1]{Institute for Theoretical Physics, Otto-von-Guericke University
Magdeburg, P.O.B. 4120, 39016 Magdeburg, Germany}
\address[cardiff]{School of Dentistry, Cardiff University, Cardiff  CF14 4XY, Wales, UK}
\address[magdeburg2]{Computing Center, Otto-von-Guericke University
Magdeburg, P.O.B. 4120, 39016 Magdeburg, Germany}

\begin{abstract}
We use the coupled cluster method (CCM) to study the
ground-state properties and lowest-lying triplet excited state of the
spin-half {\it XXZ} antiferromagnet on the square lattice.  The CCM is
applied to it to high orders of approximation by using an efficient computer
code that has been written by us and which has been implemented to run
on massively parallelized computer platforms. We are able therefore to
present precise data for the basic quantities of this model over a
wide range of values for the anisotropy parameter $\Delta$ in the
range $-1 \leq \Delta < \infty$ of interest, including both the easy-plane $(-1 < \Delta < 1)$ and easy-axis $(\Delta > 1)$ regimes, where
$\Delta \rightarrow \infty$ represents the Ising limit. We present
results for the ground-state energy, the sublattice magnetization, the
zero-field transverse magnetic susceptibility, the spin stiffness, and
the triplet spin gap. Our results provide a useful yardstick against
which other approximate methods and/or experimental studies of
relevant antiferromagnetic square-lattice compounds may now compare
their own results.  We also focus particular attention on the
behaviour of these parameters for the easy-axis system in the vicinity of the isotropic
Heisenberg point ($\Delta = 1$), where the model undergoes a phase
transition from a gapped state (for $\Delta > 1$) to a gapless state
(for $\Delta \leq 1$), and compare our results there with those from
spin-wave theory (SWT).  Interestingly, the nature of the criticality at $\Delta=1$ for the present model with spins of spin quantum number $s=\frac{1}{2}$ that is revealed by our CCM results seems to differ qualitatively from that predicted by SWT, which becomes exact only for its near-classical large-$s$ counterpart.
\end{abstract}


\end{frontmatter}


\section{Introduction}
The antiferromagnetic {\it XXZ} model on the square lattice is an important model 
that is used to describe antiferromagnetic insulators. The Hamiltonian for this system 
is given by
\begin{equation}
\mathcal{H} ~=~ \sum_{\langle i,j \rangle}[s^{x}_i s^{x}_j + s^{y}_i s^{y}_j
+ \Delta s^z_is^z_j]~~ ,
\label{XXZ_1}
\end{equation}
where the index $i$ runs over all $N$ sites on the infinite ($N \rightarrow \infty$) square lattice and the sum on ${\langle i,j \rangle}$ 
runs over all nearest-neighbour bonds on this lattice (counting each bond 
once only). 
Each site $i$ of the lattice carries a quantum spin ${\bf s}_{i} \equiv (s^{x}_{i},s^{y}_{i},s^{z}_{i})$, with ${\bf s}^{2}_{i} = s(s+1)$, and where the spin components obey the usual SU(2) commutation relations.  We shall be interested here specifically in the case $s=\frac{1}{2}$ only.  For the classical version ($s \rightarrow \infty$) of the model, it is trivial to see that for $|\Delta| > 1$ the energy is minimized (in this easy-axis case) when the spins align in the spin-space $z$ direction, to give a ferromagnetic ground state for $\Delta < -1$ and an antiferromagnetic N\'{e}el ground state for $\Delta > 1$.  Conversely, for values $|\Delta| < 1$ in the easy-plane regime, the classical ground state is again a N\'{e}el state, but now with the spins aligned parallel or antiparallel to some arbitrary direction in the $xy$ spin plane.  The classical ground-state energy per spin, $e^{{\rm cl}}_{0} \equiv E^{{\rm cl}}_{0}/N$, is thus
\begin{equation}
e^{{\rm cl}}_{0} = \left \{ \begin{array}{ll}
-2s^{2}~~; \quad \quad |\Delta| < 1 \\
-2s^{2}|\Delta|~~; \quad |\Delta|>1~~ ,
\end{array}
\right.          \label{E_classical}
\end{equation}
for classical spins of length $s$.
Whereas the ferromagnetic state is also an eigenstate of the quantum Hamiltonian for any value of the spin quantum number $s$, this is not the case for either of the N\'{e}el states, and the role of quantum fluctuations now becomes important for finite value of $s$.

Increasing experimental effort has been expended to investigate layered quantum
magnets, and precise theoretical results for the fundamental quantities, such as the
ground-state energy, the sublattice magnetization, the spin stiffness, and the uniform 
transverse magnetic susceptibility, are therefore desirable for the antiferromagnetic {\it XXZ} 
model on the square lattice.  In particular, the spin-half {\it XXZ} antiferromagnet 
on the square lattice has attracted much attention in relation to the magnetic
properties of the parent compounds of high-temperature cuprate 
superconductors \cite{manousakis91,002}.

The properties of two-dimensional (2D) bipartite (i.e., geometrically unfrustrated) lattice quantum spin systems
may be investigated by using a variety of approximate techniques (see, e.g.,
Refs.~\cite{manousakis91,lnp04}). Foremost among these for 2D unfrustrated quantum spin systems 
are various quantum Monte Carlo (QMC) simulation methods (see, e.g., Refs. \cite{cavo,qmc1,qmc2,qmc3,Beard:1996_SqLatt,qmc4,Kim:1998_SqLatt_QMC,qmc5}). 
Other approximate techniques that may be applied in order to simulate 
the properties of 2D quantum magnets include spin-wave theory (SWT) 
\cite{swt1,swt2,Igarashi:1992_SqLatt,swt3,thirdorderswt,series,Hamer:1994_SqLatt,Stinchcombe:1971_SqLatt}, exact diagonalizations (ED) 
\cite{qmc5,schulz,ed,ED40,richter2010,lauchli2011} on small finite-sized lattices, and series expansion
(SE) methods \cite{series,Hamer:1994_SqLatt,Singh:1989}. Another versatile method of {\it ab initio} quantum many-body theory
that has been shown over the last two decades to give consistently reliable 
and accurate results for 2D quantum magnetic systems at zero temperature 
is provided by the coupled cluster method (CCM) 
\cite{ccm1,ccm2,ccm5,ccm9,ccm10,ccm12,ccm13a,ccm15,ccm17,bishop04,
ccm24,ccm26,ccm27,ccm29,ccm32,ccm34,ccm35,ccm42,richter2010,honey2011,xian2011,LPSUB2011,goetze2011,bishop2012,
wir_XXZ_2015,jiang2015,gap2015,bishop2015,trian2016,Bishop:2016_honey_grtSpins}.
In particular, the use of computer-algebraic implementations \cite{ccm12,ccm15,ccm26} 
of the CCM for spin-lattice problems has increased the accuracy of the method greatly. 
It has been demonstrated conclusively in a series of recent studies (see, e.g.,
\cite{ccm24,ccm32,ccm34,ccm35,ccm42,richter2010,honey2011,goetze2011,bishop2012,wir_XXZ_2015})
that the CCM gives reliable results even in the vicinity of quantum phase transition points
for a host of quantum magnetic systems. Hence, the CCM applied to high orders of 
approximation is a good choice in order  to provide accurate results for 2D quantum 
magnetic systems. In this paper we present CCM results for the ground-state energy, 
the sublattice magnetization, the zero-field, uniform transverse magnetic susceptibility, the spin stiffness, and the 
spin gap over a wide range of values of the anisotropy parameter $\Delta$ for the Hamiltonian
given in Eq. (\ref{XXZ_1}).

We start with a brief description of the CCM  formalism in Sec.~\ref{method}, 
and then we go on to describe the application of the method to the spin-half 
{\it XXZ} model on the square lattice in Sec.~\ref{method_specific}. We present 
our results in Sec.~\ref{results}, where we also provide a discussion of their implications. 
All results are presented in graphical and tabular formats in order to provide 
a straightforward quantitative ``reference'' data set, against which results from other approximate 
methods or from experiment for relevant magnetic materials may be compared.  We conclude with a summary and discussion in Sec.\ \ref{summ_sec}.

\section{Method}
\label{method}

The details of both the fundamental and practical aspects involved in 
applying the high-order CCM formalism to lattice quantum spin systems are given, e.g., in Refs.
\cite{ccm2,ccm12,ccm13a,ccm15,bishop04,ccm26,ccm27,ccm42}. 
For the sake of brevity, we outline here only some important
features of the CCM.
First we mention that the CCM provides results 
in the infinite-lattice limit $N \rightarrow \infty$ from the outset, since it obeys the important Goldstone linked-cluster theorem at any level of approximate implementation. 
The ket and bra ground-state eigenvectors, $|\Psi\rangle$ and
$\langle\tilde{\Psi}|$, 
are parametrized within the single-reference CCM as follows   
\begin{eqnarray}\label{ket-bra_eigenvect}
|\Psi\rangle = {\rm e}^S |\Phi\rangle \; &;&  
\;\;\; S=\sum_{I \neq 0} {\cal S}_I C_I^{+}  \nonumber \; , \\ 
\langle\tilde{\Psi}| = \langle\Phi| \tilde{S} {\rm e}^{-S} \; &;& 
\;\;\; \tilde{S} =1 + \sum_{I \neq 0} \tilde{{\cal S}}_I C_I^{-} \; ,  
\label{eq2} 
\end{eqnarray} 
where $|\Phi\rangle$ is 
a suitably chosen single normalized
model or reference state. The ground-state 
ket- and bra-state Schr\"odinger  equations for a general Hamiltonian $H$ are given by 
$H |\Psi\rangle = E_0 |\Psi\rangle $ and
$\langle\tilde{\Psi}| H = E_0 \langle\tilde{\Psi}|$.  State normalizations are chosen so that $\langle\tilde{\Psi}|\Psi\rangle=\langle\Phi|\Psi\rangle=\langle\Phi|\Phi\rangle=1$.
The reference state $|\Phi\rangle$ is required to have the 
property of being a cyclic vector with respect to two well-defined Abelian 
subalgebras of {\it multi-configurational} creation operators $\{C_I^{+}\}$ 
and their Hermitian-adjoint destruction counterparts $\{ C_I^{-} \equiv 
(C_I^{+})^\dagger \}$, such that $\langle\Phi|C^{+}_{I}=0=C^{-}_{I}|\Phi\rangle,\;\forall I \neq 0$. These conditions ensure the automatic fulfillment of the above normalization conditions. The set-index $I$ denotes here a set of single-spin configurations, and the states $C^{+}_{I}|\Phi\rangle$ span the Hilbert space.  By definition, $C^{+}_{0}\equiv 1$, the identity operator.  
The correlation coefficients ${\cal S}_I$ are calculated by minimizing the 
ground-state energy expectation value functional $\bar H  =  \langle\tilde{\Psi}|H|{\Psi} 
\rangle=\bar H[{\cal S}_{I},\tilde{\cal S}_{I}]$ with respect to $\tilde{\cal S}_I$, thus leading to a coupled set 
of ket-state equations given by $\langle \Phi|C_I^- {\rm e}^{-S} H{\rm e}^S |\Phi \rangle = 0$, $\forall I \ne 0$. 
The correlation coefficients $\tilde {\cal S}_I$ are similarly found by minimizing $\bar H$ 
with respect to ${\cal S}_I$, thus leading to $\langle \Phi | {\tilde S}\left(
{\rm e}^{-S} H{\rm e}^S -E_{0}\right )C_I^+ |\Phi \rangle = 0$, $\forall I \ne 0$.
An equivalent form of this latter equation is given by 
$\langle\Phi|\tilde{S} {\rm e}^{-S} [H,C_I^{+}] {\rm e}^S|\Phi\rangle 
= 0$.

An excited state $|\Psi_e\rangle$ is parametrized within the CCM by applying an 
excitation operator $X^e$ linearly to the ground state $|\Psi\rangle$,  such that
\begin{equation}
 \label{eqngap_1}
|\Psi_e\rangle = X^e{\rm e}^S|\Phi\rangle~~ ; ~~  X^e = \sum_{I\not=0} \mathcal{X}_I^eC_I^+~~.
\end{equation}
From the Schr\"odinger equation, $H|\Psi_e\rangle = E_e|\Psi_e\rangle$, it follows that
\begin{equation}\label{eqngap2}
{\rm e}^{-S}[H,X^e] {\rm e}^S|\Phi\rangle = \varepsilon X^e|\Phi\rangle~~ ,   
\end{equation}
where $\varepsilon \equiv (E_e-E_0)$ is the excitation energy. We now project Eq.\ (\ref{eqngap2}) on the left with the state $\langle \Phi|C_I^-$, and use that the states labeled by the indices $I$ are, as usual, orthormalized, $\langle\Phi|C^{-}_{I}C^{+}_{J}|\Phi\rangle=\delta(I,J)$, to yield the generalized set of eigenvalue equations
\begin{equation}\label{eqngap3}
\langle\Phi|C_I^-{\rm e}^{-S}[H,X^e]{\rm e}^S|\Phi\rangle=\varepsilon \mathcal{X}_I^e  ~~,
\end{equation}
which we solve in order to obtain $\varepsilon$. 
In the present case we will be interested specifically in the case
when $|\Psi_{e}\rangle$ is the lowest-lying triplet excited state,
above the spin-singlet ground state $|\Psi\rangle$, and $\varepsilon$
is hence the (triplet) spin gap.

The CCM formalism is exact in the limit of inclusion of all possible multi-spin clusters 
within the ground- and excited-state operators [i.e., by inclusion of all multi-spin configurations $I$ in the sums in Eqs.\ (\ref{ket-bra_eigenvect}) and (\ref{eqngap_1})], although this is usually impossible to 
achieve practically. The so-called LSUB$m$ approximation 
scheme is used here for both the ground and excited states. This approximation
scheme uses all multi-spin correlations over all distinct cluster locales on the lattice defined 
by $m$ or fewer contiguous sites. Such locales (or lattice animals) of size $m$ are said to be contiguous if every site in the cluster is nearest-neighbour to at least one other. We select equivalent levels of LSUB$m$ 
approximation for both the ground and excited states. However, we remark that for our calculation of the (triplet) spin gap 
the choice of clusters for the lowest-lying (triplet) excited state is different 
from those for the ground state because we know that the ground state lies in the 
$s_T^z (\equiv \sum^{N}_{i=1} s_i^z) = 0$ subspace, whereas the lowest-lying triplet excited state 
in terms of energy must have $s_T^z = \pm 1$. Hence, we only 
use configurations in the excited-state operator $X^e$ that change the total spin 
by one. We find that  the number of configurations for the excited state is 
larger than for the ground state at all levels of LSUB$m$ approximation. 
The number of terms in the corresponding equation systems is correspondingly larger, 
and so the calculation of the excited state is more difficult computationally than
that of the ground state.

The LSUB$m$ approximation scheme allows the systematic analysis  of CCM data 
as a function of the level of approximation $m$, without any further approximations being made.  We extrapolate the 
individual LSUB$m$ data to the limit $m \rightarrow \infty$ in order to form accurate 
estimates of all expectation values. The general form for extrapolating LSUB$m$ results 
in the limit $m \to\infty$ is given by $A(m)=A_0+A_1(1/m)^{\nu_1}+A_2(1/m)^{\nu_2}$, 
where the (fixed) leading exponents $\nu_1$ and $\nu_2$ ($> \nu_1$) may be different
for the different quantities to be extrapolated (and see Sec.~\ref{method_specific}
for details).  Finally, we note that at any LSUB$m$ level of approximation the CCM exactly fulfills both the Goldstone linked-cluster theorem and the very important Hellmann-Feynman theorem.

\section{The CCM applied to the {\it XXZ} Model}
\label{method_specific}
We recall that the spin-half {\it XXZ} antiferromagnetic model on the
square lattice with nearest-neighbour interactions is given by
Eq.~(\ref{XXZ_1}).  Here we use the quasiclassical $z$-aligned N\'eel
state as the model state $|\Phi\rangle$ for values of the anisotropy
parameter in the range $\Delta \ge 1$, whereas we use a N\'eel state
aligned in the $xy$ plane for $-1 \le \Delta \le 1$. Both reference
states give identical results for the rotationally invariant model at
$\Delta =1$.  It is convenient to carry out a transformation of the
local spin axes on each site such that all spins in each reference
state align along the negative $z$ axis.  A complete set of multi-spin
creation operators may then be formed with respect to every model
state, and we note that this set of {\it multi-configurational}
creation operators with respect to the rotated coordinate frame is
defined by
$\{C_I^+= s_{i_1}^+s_{i_2}^+\cdots s_{i_n}^+\; ; \;
n=1,2,\ldots,2sN\}$,
where $s^{\pm}_{k}\equiv s^{x}_{k} \pm is^{y}_{k}$.  As we are
henceforth interested only in the case $s=\frac{1}{2}$, we note that
no site index $i_{k}$ contained in any retained cluster index $I$ may
appear more than once.  In the LSUB$m$ approximation for the present
$s=\frac{1}{2}$ case therefore, we retain in the sums over multi-spin
configurations $I$ in Eqs.\ (\ref{eq2}) and (\ref{eqngap_1}) only
those terms involving the set-indices
$I=\{i_{1},i_{2},\ldots,i_{n}\; ; \; n=1,2,\ldots,N\}$ where
$n \leq m$, and where each site index $i_{k} \in I$ is
nearest-neighbour to at least one other site index $i_{l} \in I$.

For the $z$-aligned N\'eel model state we perform a rotation of all 
``up-pointing'' spins (say, on the $B$ sublattice) by 180$^{\rm o}$ 
about the $y$-axis. The transformation of the local axes of the $B$-sublattice 
spins is given by
\begin{equation}
s^x_j ~\rightarrow ~ -s^x_j~, ~ s^y_j ~\rightarrow ~ s^y_j~, ~ s^z_j ~\rightarrow ~ -s^z_j~~ .
\end{equation}
The local spin axes of the  ``down-pointing'' spins (say, on the $A$ sublattice) do not need to be rotated. 
The Hamiltonian of Eq. (\ref{XXZ_1}) within the rotated coordinate frame is  given by
\begin{equation}
\mathcal{H}=-\frac{1}{2} \sum_{\langle i,j \rangle} [s^+_is^+_j+s^-_is^-_j
+2\Delta s^z_is^z_j]~~,
\label{XXZ_2}
\end{equation}
for the N\'eel model state with spins aligned in the $z$ direction and with respect to the rotated spin axes.

We use the N\'eel state with spins aligned along the  $x$ axis as the model 
state in the regime given by $-1 \le \Delta \le 1$.
We rotate the axes of the left-pointing spins (i.e., those pointing along the negative $x$-direction on, say, sublattice $A$) by 90$^\circ $ about the $y$ axis, whereas we 
rotate the axes of the right-pointing (i.e., those pointing along the positive $x$-direction on, say, sublattice $B$) spins by 270$^\circ$ about the $y$ axis. The corresponding  transformation of the local spin axes on sublattice $A$ is given by 
\begin{equation}
s^x_i~ \rightarrow~ -s^z_i~, ~ s^y_i ~ \rightarrow ~  s^y_i ~ , ~
s^z_i ~ \rightarrow ~ s^x_i~~ ;  \label{eq:leftRotation}
\end{equation}
and the corresponding transformation of the local spin axes on sublattice $B$ is given by 
\begin{equation}
s^x_j~ \rightarrow~ s^z_j~,~ s^y_j ~ \rightarrow ~  s^y_j~ , ~
 s^z_j~ \rightarrow ~ -s^x_j~~ .  \label{eq:rightRotation}
\end{equation}
The Hamiltonian of Eq. (\ref{XXZ_1}) is then given by
\begin{equation}
\mathcal{H} ~=~ -\frac 14 \sum_{\langle i,j \rangle}  \; \left[ \; 
(\Delta+1)(s_i^+s_j^+ + s_i^-s_j^-) 
 + (\Delta -1)(s_i^+ s_j^- + s_i^{-} s_j^+ )
+4s_i^z s_j^z  \; \right] ~~ ,
\label{newHamiltonian}
\end{equation}
for the N\'eel model state with spins aligned in the $xy$ plane and with respect to the rotated spin axes.

We are able to evaluate ground-state expectation 
values of arbitrary operators once the values 
for the bra- and ket-state correlation coefficients, $\tilde{{\cal S}}_{I}$ and ${\cal S}_{I}$ respectively, have 
been determined (at a given level of approximation), as described in Sec.\ \ref{method}. The ground-state energy per spin is given, uniquely, in terms of the coefficients $\{{\cal S}_{I}\}$ alone, by
\begin{equation} 
e_0\equiv\frac{E_0}{N}=\frac{1}{N}\langle\Phi|e^{-S}He^S|\Phi\rangle ~~.
\label{E}
\end{equation}
The sublattice magnetization is given in terms of the rotated spin coordinates for {\it both} model states  by
\begin{equation}
M ~=~ - \frac 1{N} \langle \tilde{\Psi}| \sum_{i=1}^N s_i^z | \Psi \rangle
~=~  - \frac 1N  \langle \Phi | \tilde{S} e^{-S} (\sum_{i=1}^N s_i^z) e^S | \Phi \rangle
~~.
\label{sublatticemagnetisation}
\end{equation}
The classical ($s \rightarrow \infty$) version of the model has a sublattice magnetization $M_{{\rm cl}}=s$ for each of the ground-state phases.  For the quantum version, when $s$ takes a finite value, $M$ remains equal to its classical value only in the ferromagnetic phase.  For each of the two N\'{e}el phases one expects that quantum fluctuations will reduce the value of $M$ below its classical counterpart.
 
The transverse uniform magnetic susceptibility may be calculated within the CCM by using the method outlined in Refs.~\cite{ccm42,trian2016} for the square- and triangular-lattice Heisenberg antiferromagnet. However, it is useful to note here briefly that we add an appropriate transverse magnetic field term 
to the Hamiltonian of Eq.\ (\ref{XXZ_1}), 
namely: $-\lambda \sum_i s^x_i$ for the $z$-aligned N\'eel 
reference state ($\Delta \ge 1$); or, $-\lambda \sum_i s^z_i$ for 
the $x$-aligned N\'eel reference  state ($|\Delta| \le 1$), both in units where the gyromagnetic ratio $g\mu_{B}/\hbar=1$. 
Spins are now allowed to cant at an angle, 
and this angle tends to zero in the limit $\lambda \rightarrow 0$. 
The precise nature of the canted model states and the solution of 
the associated CCM problem is described in detail in  Refs.~\cite{ccm42,trian2016}. 
The uniform transverse magnetic susceptibility is then defined as usual by the relation
\begin{equation} \label{susc}
\chi(\lambda) =
- \frac{1}{N}\frac{d^2 E_0}{d\lambda^2} ~~ ,
\end{equation}
where we now calculate the ground-state energy, $E_{0}=E_{0}(\lambda)$, in the presence of the applied magnetic field.  The zero-field susceptibility, $\chi \equiv \chi(0)$, may be calculated from the small-$\lambda$ expansion,
\begin{equation}
\frac {E_0(\lambda)}N =\frac {E_0(\lambda=0)}N -\frac 12 \chi \lambda^2 + {\cal O} (\lambda^4)~~ .
\label{chi_expansion}
\end{equation}
For the classical version of the model it is easy to show that $\chi$ takes the same value,
\begin{equation}
\chi_{{\rm cl}} = \frac{1}{4(1+\Delta)}~~ ; ~~ -1 < \Delta < \infty~~ ,  \label{chi_classical_eq}
\end{equation}
in both ground-state N\'{e}el phases, independent of the length $s$ of the classical spins.

The calculation of the spin stiffness $\rho_s$ using the CCM is described in Refs.~\cite{ccm27,ccm29,ccm35,bishop2015,trian2016}. The spin stiffness measures the increase 
in the amount of energy for 
a magnetically long-range ordered system when a helical ``twist'' of magnitude 
$\theta$ per unit length is imposed on the spins, in a given direction. In this case the ground-state energy  per spin is given by
\begin{equation}
\frac {E_0(\theta)}N =\frac {E_0(\theta=0)}N +\frac 12 \rho_s \theta^2 + {\cal O} (\theta^4)~~ ,
\label{stiffness}
\end{equation}
where $E_0(\theta)$ is the ground-state energy as a function of the imposed twist, (see, e.g., Refs.~\cite{schulz1995,lhuillier1995,trumper1998} for details).
Again, we use a rotation of the local spin at site $i$ by an appropriate angle $\delta_i$ such that the local spin axes
for the now helical reference state appear mathematically to align along the (negative) $z$ axis (for details see Refs.~\cite{ccm27,ccm29}). The helical state lies in the $xy$ plane for $\Delta \le 1$, and is thus well-defined to give a unique determination of $\rho_{s}$. For the classical version of the model it is simple to show that $\rho_{s}$ takes the classical value,
\begin{equation}
\rho^{{\rm cl}}_{s} = s^{2}~~ ; ~~ -1 < \Delta < 1~~ ,  \label{rho_classical_eq}
\end{equation}
for classical spins of length $s$, in units where the nearest-neighbour spacing on the square lattice has been set to unity.  By contrast, the spin stiffness is ill-defined for $\Delta > 1$ because the helical state lies in the $xz$ plane. The easy-axis anisotropy therefore adds an energy contribution proportional to $\cos (\delta_i)$, and so the energy depends on the individual angles $\delta_i$ relative to the easy axis.

As already outlined briefly in Sec.~\ref{method}, as a final step we need to extrapolate our LSUB$m$ estimates for all physical quantities to the limit $m \rightarrow \infty$ where the method becomes exact.  Although exactly provable rules are not known for these extrapolations, robust empirical rules do exist, and these rules have successfully been tested for a wide range of quantum magnetic systems 
\cite{ccm15,bishop04,ccm27,ccm29,ccm42,gap2015,trian2016}.
We use the ``standard'' rules in order to extrapolate all expectation values, namely: the ground-state energy per spin $e_0 \equiv E_0/N$ using $e_0(m) = a_0 + a_1/m^2 + a_2/m^4$; the sublattice magnetization using $M(m)=b_0+b_1/m+b_2/m^{2}$; 
the zero-field, uniform transverse magnetic susceptibility using $\chi(m)=c_0+c_1/m+c_2/m^2$;
the spin stiffness using $\rho_s(m)=d_0+d_1/m+d_2/m^2$;
and the spin gap using $\varepsilon(m)=f_0+f_1/m+f_2/m^2$.

The numbers, $N_{f}=N_{f}(m)$, of distinct (fundamental) configurations $I$ that are retained in the summations for both the ground state in Eq.\ (\ref{eq2}) and the excited state in Eq.\ (\ref{eqngap_1}) at a given LSUB$m$ level of approximation are reduced by utilizing the space- and point group symmetries of the Hamiltonian and the model state, together with any conservation laws that pertain to both the Hamiltonian and the specific model state being used (viz., specifically here for $s^{z}_{T}$). 
We are able to compute data up to the order LSUB12 for the ground-state energy $e_0$, the sublattice magnetization $M$, and the spin gap $\varepsilon$ using the high-order CCM code \cite{thecode}. The maximum number of fundamental ground-state
configurations used in our calculations is $N_f(12) = 4
\; 248 \; 225$, and this calculation was carried out for the planar N\'{e}el model ground state at the LSUB12 level of approximation. The solution of the LSUB$m$ 
equations is more challenging for the susceptibility $\chi$  and the spin stiffness $\rho_s$ because less symmetries can be used in these cases. As a result we can calculate the magnetic
susceptibility and the spin stiffness only up to the LSUB10 level of approximation. Finally, we extrapolate our LSUB$m$ results 
for the ground-state energy $e_0$, the sublattice magnetization $M$, and the spin gap $\varepsilon$ by using data for $m=\{4,6,8,10\}$ and then separately also by using data for $m=\{4,6,8,10,12\}$. In this manner, we provide two sets of extrapolated values for $e_0$, $M$, and $\varepsilon$. By comparing these two sets of estimates, we obtain an estimate of the precision of these extrapolated quantities. We refer to extrapolated results using LSUB$m$ results for $m=\{4,6,8,10\}$ and $m=\{4,6,8,10,12\}$ as LSUB$\infty(1)$ and LSUB$\infty(2)$, respectively. 

\begin{figure}[!t]
\center
\includegraphics[width=0.9\textwidth]{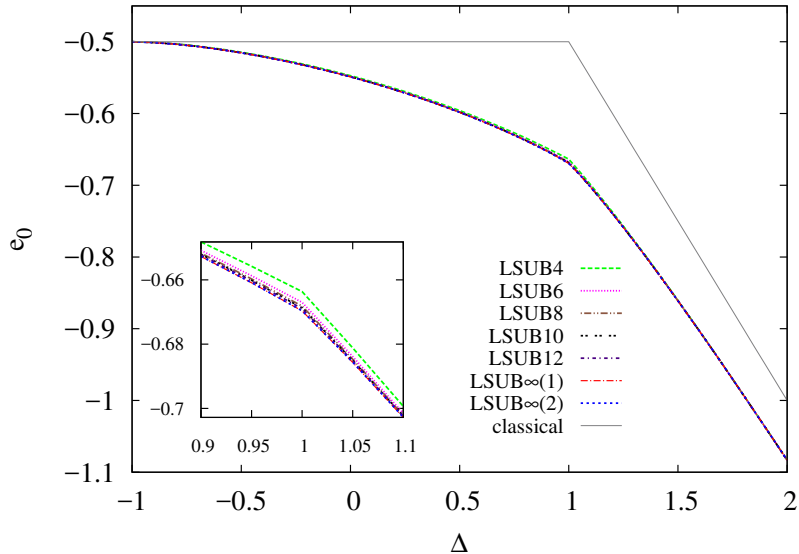}
\caption{\label{XXZ_fig1} The ground-state energy per site $e_{0} \equiv E_{0}/N$ for the spin-$\frac{1}{2}$ $XXZ$ antiferromagnet on the square lattice as a function of the anisotropy parameter $\Delta$.   The CCM model states used are N\'{e}el states aligned respectively in the $x$ direction for $-1 \leq \Delta \leq 1$ and in the $z$ direction for $\Delta \geq 1$.  We show CCM LSUB$m$ results for $m=4,6,8,10,12$, together with extrapolated LSUB$\infty(1)$ and LSUB$\infty(2)$ results based on the data sets $m=\{4,6,8,10\}$ and $m=\{4,6,8,10,12\}$, respectively.  Note that in the main
panel the lines 
practically coincide.  The inset shows the region near $\Delta=1$ in more detail.  We also show the corresponding classical result from Eq.\ (\ref{E_classical}) with $s=\frac{1}{2}$.}
\end{figure}

We remark that the results presented in this article are carried out to much higher levels of LSUB$m$ approximation than those presented in previous CCM investigations of the {\it XXZ} model \cite{ccm5,ccm9,ccm10,xian2011}, where the highest order of approximation was the LSUB8 approximation. The consequent accuracy of our results is thus significantly higher than those presented in Refs.~\cite{ccm5,ccm9,ccm10,xian2011}. Moreover, a systematic study of the magnetic susceptibility and the spin stiffness of the {\it XXZ} model was not presented in these earlier studies.

\section{Results}
\label{results}
\begin{figure}[!t]
\center
\includegraphics[width=0.9\textwidth]{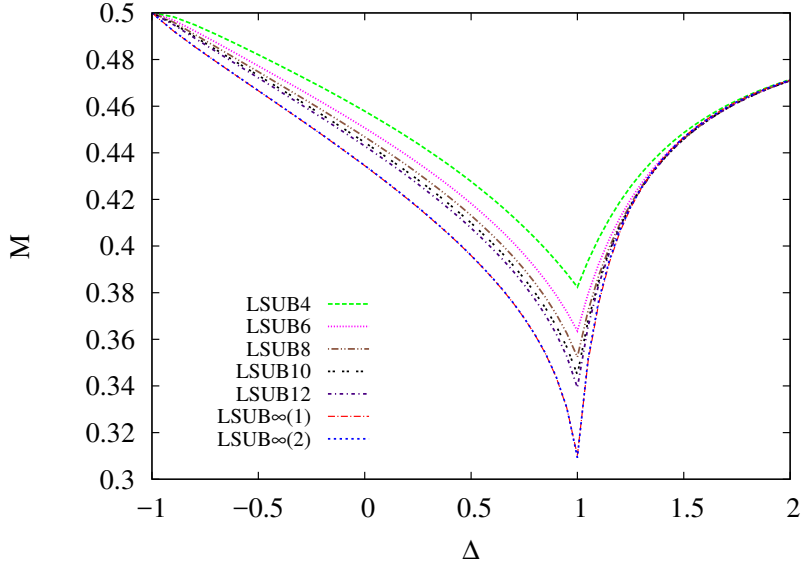}
\caption{\label{XXZ_fig2} The sublattice magnetization $M$ for the spin-$\frac{1}{2}$ $XXZ$ antiferromagnet on the square lattice as a function of the anisotropy parameter $\Delta$.   The CCM model states used are N\'{e}el states aligned respectively in the $x$ direction for $-1 \leq \Delta \leq 1$ and in the $z$ direction for $\Delta \geq 1$.  We show CCM LSUB$m$ results for $m=4,6,8,10,12$, together with extrapolated LSUB$\infty(1)$ and LSUB$\infty(2)$ results based on the data sets $m=\{4,6,8,10\}$ and $m=\{4,6,8,10,12\}$, respectively.}
\end{figure}

We first show in Figs.\ \ref{XXZ_fig1} and \ref{XXZ_fig2} our CCM
results for the ground-state energy per site, $e_{0} \equiv E_{0}/N$,
and the ground-state sublattice magnetization $M$ pertaining to the spin-$\frac{1}{2}$ Hamiltonian of Eq.\ (\ref{XXZ_1}) on the square lattice.  In both figures we
show results obtained in LSUB$m$ approximations with
$m=4,6,8,10,12$, using as CCM model states an $x$-aligned N\'{e}el
state in the range $-1 < \Delta < 1$ and a $z$-aligned N\'{e}el state
in the range $\Delta > 1$ of the anisotropy parameter.  

We note that these model states provide exact ground states of the
Hamiltonian of Eq.\ (\ref{XXZ_1}) in the respective limits
$\Delta = -1$ and $\Delta \rightarrow \infty$ (the Ising limit).
Thus, exact results for all ground-state quantities are achieved for
these two limiting cases at all LSUB$m$ levels of approximation (viz.,
$e_{0} = -\frac{1}{2}$ and $M = \frac{1}{2}$ at $\Delta = -1$, and
$e_{0} = -\frac{1}{2}\Delta$ and $M = \frac{1}{2}$ at
$\Delta = \infty$).  In each of Figs.\ \ref{XXZ_fig1} and
\ref{XXZ_fig2} we also show two sets of extrapolated (LSUB$\infty$)
results, based on the respective schemes described in Sec.\
\ref{method_specific}, and using the two appropriate LSUB$m$ input
data sets with $m=\{4,6,8,10\}$ and $m=\{4,6,8,10,12\}$.

Figure \ref{XXZ_fig1} shows that our CCM results for the
ground-state energy converge very rapidly as the order $m$ of the
LSUB$m$ approximation is increased towards the exact
($m \rightarrow \infty$) limit.  Indeed, both the raw LSUB$m$ results
and the two LSUB$\infty$ extrapolations, based on the two different
input LSUB$m$ data sets as described above, are difficult to resolve
by eye in the main panel of Fig.\ \ref{XXZ_fig1}, which shows
the high accuracy achieved within the CCM LSUB$m$ framework for the
energy.  The first-order transition at $\Delta = 1$ between the two
N\'{e}el forms of long-range order (viz., that aligned in the $xy$
plane for $|\Delta| < 1$ and that aligned along the $z$ axis for
$\Delta >1$) is clearly visible in the $e_{0}=e_{0}(\Delta)$ curves
shown in Fig.\ \ref{XXZ_fig1}.  The inset to Fig.\ \ref{XXZ_fig1}
presents the results near the critical point at $\Delta = 1$ in more
detail.

An estimate of the accuracy of our extrapolated results can be
obtained by a comparison of the two different extrapolation schemes, LSUB$\infty(1)$ and LSUB$\infty(2)$.  For example, our LSUB$\infty$
results at the isotropic Heisenberg ($XXX$) point ($\Delta = 1$) are
$e_{0} = -0.66966$ using the LSUB$m$ data set $m=\{4,6,8,10\}$ and
$e_{0} = -0.66964$ using the LSUB$m$ data set $m=\{4,6,8,10,12\}$.
Corresponding results at the isotropic $XY$ ($\equiv XX$) point
($\Delta = 0$) are $e_{0} = -0.54890$ using the LSUB$m$ data set
$m=\{4,6,8,10\}$ and $e_{0} = -0.54888$ using the LSUB$m$ data set
$m=\{4,6,8,10,12\}$.  It is clear that the results for the ground-state energy are very insensitive to the extrapolation procedure.  We estimate that over the whole range of values of $\Delta$, our accuracy is better than 1 part in $10^{4}$.

Our corresponding results for the sublattice magnetization $M$ are
shown in Fig.\ \ref{XXZ_fig2}.  As is fully to be expected the results
for the order parameter are both more strongly dependent on the order
$m$ of the LSUB$m$ approximation and converge more slowly as
$m \rightarrow \infty$.  Just as for the ground-state energy the two
LSUB$\infty$ extrapolations, based on LSUB$m$ results with
$m=\{4,6,8,10\}$ and $m=\{4,6,8,10,12\}$ respectively, are almost
indiscernible in Fig.\ \ref{XXZ_fig2}.  The maximum difference in the
two extrapolations is at the isotropic Heisenberg point, $\Delta=1$,
where from Fig.\ \ref{XXZ_fig2} we see that the effect of quantum
fluctuations is largest at reducing the order parameter from its
classical value $M_{{\rm cl}}=\frac{1}{2}$.  Thus, our LSUB$\infty$
results at the isotropic Heisenberg ($XXX$) point ($\Delta=1$) are
$M=0.31024$ using the LSUB$m$ data set $m=\{4,6,8,10\}$ and
$M=0.30931$ using LSUB$m$ data set $m=\{4,6,8,10,12\}$.  The relative
error between the two results is thus of the order of 3 parts in
$10^{3}$.  By comparison, the corresponding LSUB$\infty$ results at
the isotropic $XY (\equiv XX)$ point ($\Delta=0$) are $M=0.43446$
using the LSUB$m$ data set with $m=\{4,6,8,10\}$ and $M=0.43458$ using
the LSUB$m$ data set with $m=\{4,6,8,10,12\}$.  The relative error
between the two extrapolations is now only of the order of 3 parts in
$10^{4}$.

Our CCM results shown in Fig.\ \ref{XXZ_fig2} imply that the
classical Ising limit, $M_{{\rm cl}}=\frac{1}{2}$, is approached
rather rapidly as the anisotropy parameter $\Delta$ is increased.  For
example, even at a value $\Delta=2$, the order parameter $M$ already
attains a value of about $94\%$ of the classical value, and for all
values $\Delta \geq 5$ the value of $M$ is greater than $99\%$ of the
classical limit.

It is interesting to compare our results for the spin-$\frac{1}{2}$ model in the vicinity of the isotropic Heisenberg point, $\Delta = 1$, with those of SWT, which are applicable in the high-spin ($s \rightarrow \infty$) classical limit.  Thus, SWT predicts \cite{series,swt2,Stinchcombe:1971_SqLatt} that in the vicinity of the isotropic point $\Delta = 1$ all of the physical ground-state parameters are analytic functions of the quantity $(1-\Delta^{-2})^{1/2}$ for $\Delta > 1$.  Hence SWT predicts that any physical parameter $R$ of the model that pertains to the scaled Hamiltonian $\mathcal{H}/\Delta$ of Eq.\ (\ref{XXZ_1}) would have an expansion $R=\sum^{\infty}_{n=0}r_{n}(1-\Delta^{-2})^{n/2}$ in the region $\Delta > 1$.  In particular, the ground-state energy and order parameter are predicted (and see, e.g., Ref.\ \cite{series}) to behave as
\begin{equation}
\frac{E^{\rm SWT}_{0}}{N\Delta}=\epsilon_{0}+\epsilon_{2}(1-\Delta^{-2})+\epsilon_{3}(1-\Delta^{-2})^{\frac{3}{2}}+\cdots ~~ ,  \label{E-eq_SWT}
\end{equation}
\begin{equation}
M^{\rm SWT}=\mu_{0}+\mu_{1}(1-\Delta^{-2})^{1/2}+\mu_{2}(1-\Delta^{-2})+\cdots ~~ .  \label{M-eq_SWT}
\end{equation}
Naively, one might expect that the phenomenology of SWT, which is
strictly valid only in the $s \rightarrow \infty$ limit, including
these functional forms, could remain correct for finite values of $s$,
at least so long as long-range antiferromagnetic N\'{e}el order
persists (i.e., $\mu_{0} > 0$) at $\Delta=1$ in the quantum model.
That is certainly the case here, since we find $\mu_{0} \approx 0.31$ at $\Delta=1$.
Thus, it is tempting to hypothesize that since the SWT singularities
in the physical parameters near $\Delta = 1$ [i.e., the odd powers in
$(1-\Delta^{-2})^{1/2}$ in the expansions] are caused by the Goldstone
modes and not by critical fluctuations, the associated leading
critical exponents for finite values of the spin quantum number $s$
should therefore be the same as predicted by SWT, even for the $s=\frac{1}{2}$
case considered here.

\begin{figure}[!t]
\center
\includegraphics[width=0.8\textwidth]{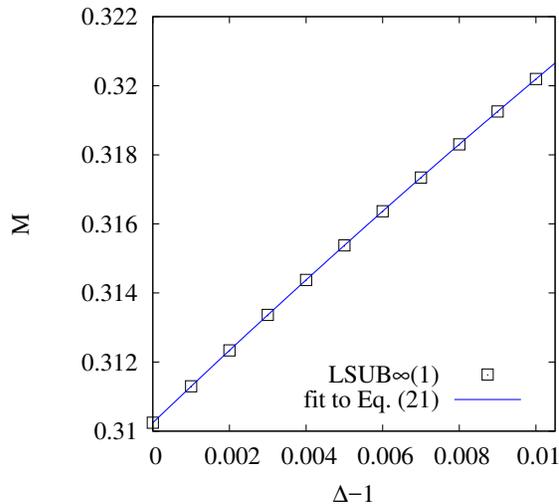}
\caption{\label{fit-M-near-heisenberg} The staggered magnetization $M$ for the spin-$\frac{1}{2}$ $XXZ$ antiferromagnet with anisotropy parameter $\Delta$ on the square lattice, plotted as a function of $(\Delta-1)$ in the vicinity of the Heisenberg point ($\Delta=1$).  The extrapolated LSUB$\infty(1)$ data points obtained from our CCM LSUB$m$ results based on the $z$-aligned N\'{e}el state as model state with $m=4,6,8,10$ are shown by open squares ($\Box$), and the solid line is the best fit to them of the form of Eq.\ (\ref{M_fit}).}
\end{figure}

In order to test this hypothesis we have carefully examined our CCM
results for the magnetic order parameter $M$ in the narrow range
$1 \leq \Delta \leq 1.01$.  We show in Fig.\
\ref{fit-M-near-heisenberg} our LSUB$\infty(1)$ extrapolations based on
the LSUB$m$ data set $m=\{4,6,8,10\}$ in this range, plotted as a
function of the parameter $(\Delta -1)$.
In order to find the leading (critical) exponent we have fitted the
data to the totally unbiased form $M=n_{0}+n_{1}(\Delta-1)^{\nu}$,
where each of the parameters $n_{0}$, $n_{1}$ and $\nu$ is fitted.
The best fit to the data points shown in Fig.\ \ref{fit-M-near-heisenberg} is obtained with $n_{0}=0.31022 \pm 0.00002$,
$n_{1}=0.826 \pm 0.017$ and $\nu=0.959 \pm 0.004$.  Since the leading exponent
takes the value $\nu \approx 1$, we thus attempt a fit of the form
\begin{equation}
M=m_{0}+m_{1}(\Delta-1)+m_{2}(\Delta-1)^{2} ~~ ,  \label{M_fit}
\end{equation}
with $m_{0}$ fixed at the value $m_{0}=0.310243$ appropriate to the
LSUB$\infty(1)$ value for $\Delta=1$, obtained as described above using
the LSUB$m$ data set with $m=\{4,6,8,10\}$.  The best fit, shown as the solid line
in Fig.\
\ref{fit-M-near-heisenberg}, is obtained with
$m_{1}=1.0592 \pm 0.0003$ and $m_{2}=-6.42 \pm 0.04$.  Thus, perhaps
surprisingly, the SWT hypothesis is {\it not} confirmed by our
results.  The square-root cusp in $M$ that is predicted by SWT appears
to be entirely absent.  Of course it is possible that for this
$s=\frac{1}{2}$ model the parameter $\mu_{1}$ in Eq.\ (\ref{M-eq_SWT})
vanishes (or takes a very small value) accidentally.  More likely,
however, is the scenario that the series for $M$ for the
spin-$\frac{1}{2}$ model is actually analytic in $(1-\Delta^{-1})$, possibly multiplied by some additional slowly varying non-algebraic (e.g., logarithmic) term, near the
isotropic Heisenberg point, rather than in the parameter
$(1-\Delta^{-2})^{1/2}$ predicted by SWT, as is appropriate in the classical
($s \rightarrow \infty$) limit.

\begin{figure}[!t]
\center
\includegraphics[width=0.9\textwidth]{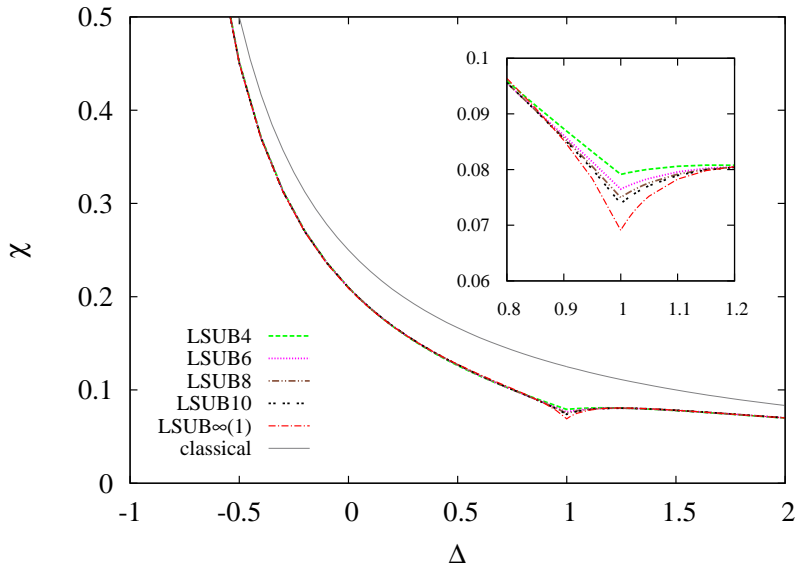}
\caption{\label{XXZ_fig3} The zero-field, uniform transverse magnetic susceptibility
$\chi$ for the spin-$\frac{1}{2}$ $XXZ$ antiferromagnet on the square lattice as a function of the anisotropy parameter $\Delta$.   The CCM model states used are canted N\'{e}el states, aligned respectively in the $x$ direction for $-1 \leq \Delta \leq 1$ and in the $z$ direction for $\Delta \geq 1$ when the external magnetic field is zero.  We show CCM LSUB$m$ results for $m=4,6,8,10$, together with the extrapolated LSUB$\infty(1)$ results based on this data set.  The inset shows the region near $\Delta=1$ in more detail.  We also show the corresponding classical result from Eq.\ (\ref{chi_classical_eq}).    
}
\end{figure}

We turn next to our results for the zero-field, uniform transverse magnetic 
susceptibility $\chi$ of the model.  Thus, we show in Fig.\
\ref{XXZ_fig3} the CCM LSUB$m$ results with $m=4,6,8,10$ and the
corresponding LSUB$\infty$ extrapolation based on this set, for the
same range of values for the anisotropy parameter,
$-1 \leq \Delta \leq 2$, as shown in Figs.\ \ref{XXZ_fig1} and
\ref{XXZ_fig2} above for the ground-state energy and sublattice
magnetization respectively.
Once again we remark that the results become exact in both limits $\Delta = -1$ and $\Delta \rightarrow \infty$ (the Ising limit).  It is clear that the LSUB$m$ sequence of results for $\chi$ converges extremely rapidly, with the curves difficult to resolve by eye over most of the range shown, except for a small region around $\Delta = 1$, where quantum fluctuations are again greatest.  The inset to Fig.\ \ref{XXZ_fig3} again presents the results near the critical point at $\Delta = 1$ in more detail.  SWT again predicts (and see, e.g., Ref.\ \cite{series}) a square-root cusp for $\chi$ near the Heisenberg point for values $\Delta > 1$, 
\begin{equation}
\Delta\chi^{{\rm SWT}} = \zeta_{0}+\zeta_{1}(1-\Delta^{-2})^{1/2}+\zeta_{2}(1-\Delta^{-2})+\cdots ~~ ,  \label{chi-eq_SWT}
\end{equation}
which appears also not to be borne out by our results in Fig.\
\ref{XXZ_fig3} for the spin-$\frac{1}{2}$ model.

\begin{figure}[!t]
\center
\includegraphics[width=0.9\textwidth]{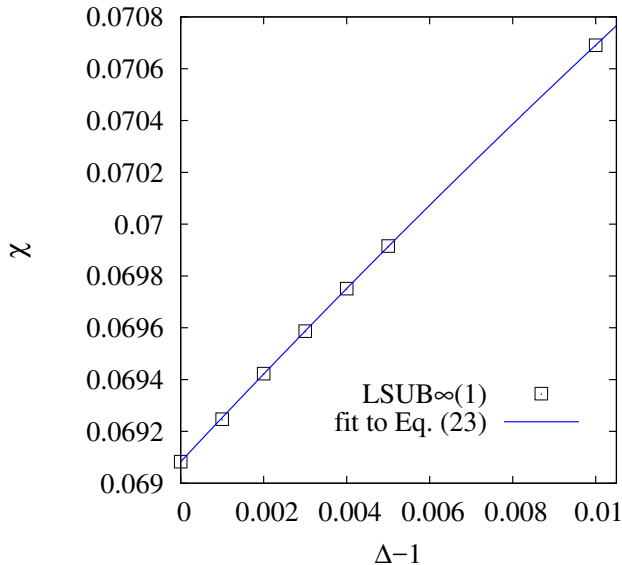}
\caption{\label{fit-chi-near-heisenberg} The zero-field, uniform transverse magnetic susceptibility
$\chi$ for the spin-$\frac{1}{2}$ $XXZ$ antiferromagnet with anisotropy parameter $\Delta$ on the square lattice, plotted as a function of $(\Delta-1)$ in the vicinity of the Heisenberg point ($\Delta=1$).  The extrapolated LSUB$\infty(1)$ data points obtained from our CCM LSUB$m$ results based on the canted N\'{e}el state (aligned in the $z$ direction when the external magnetic field is zero) as model state with $m=4,6,8,10$ are shown by open squares ($\Box$), and the solid line is the best fit to them of the form of Eq.\ (\ref{chi_fit}).}
\end{figure}

Hence, once again we show in Fig.\ \ref{fit-chi-near-heisenberg} 
our extrapolated LSUB$\infty(1)$ results for the zero-field, uniform
transverse magnetic susceptibility $\chi$ in the narrow range
$1 \leq \Delta \leq 1.01$, based on our LSUB$m$ results with
$m=4,6,8,10$.  The leading (critical) exponent $\nu$ is again
obtained by fitting the LSUB$\infty(1)$ data to the unbiased form
$\chi=y_{0}+y_{1}(\Delta-1)^{\nu}$, where each of the parameters
$y_{0}$, $y_{1}$ and $\nu$ is fitted.  The best fit to the data
points shown in Fig.\ \ref{fit-chi-near-heisenberg} is obtained with
$y_{0}=0.069078 \pm 0.000005$, $y_{1}=0.133 \pm 0.005$ and
$\nu=0.958 \pm 0.009$.  Just as for the previous fit for the staggered
magnetization $M$, the leading exponent $\nu$ again takes a value very
close to unity.  We thus attempt now a fit of the form
\begin{equation}
\chi=x_{0}+x_{1}(\Delta-1)+x_{2}(\Delta-1)^{2} ~~ ,  \label{chi_fit}
\end{equation}
with $x_{0}$ fixed at the value $x_{0}=0.069083$ appropriate to the
LSUB$\infty(1)$ value for $\Delta=1$, obtained as described previously
using the LSUB$m$ data set with $m=\{4,6,8,10\}$.  The best fit, shown
in Fig.\ \ref{fit-chi-near-heisenberg} by the solid line, is obtained
with the values $x_{1}=0.1713 \pm 0.0007$ and $x_{2}=-1.05 \pm 0.08$.

\begin{figure}[!t]
	\center
        \includegraphics[width=0.9\textwidth]{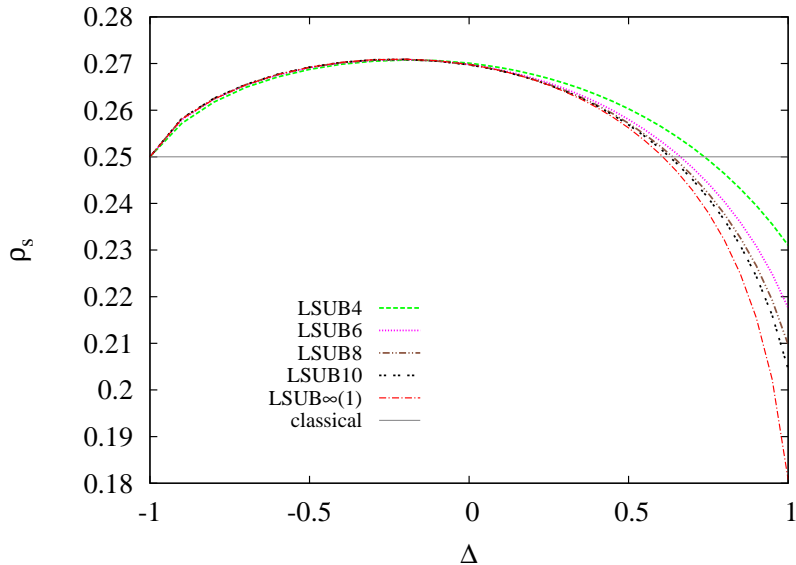}
	\caption{\label{XXZ_fig4} The spin stiffness $\rho_{s}$ for the spin-$\frac{1}{2}$ $XXZ$ antiferromagnet on the square lattice as a function of the anisotropy parameter $\Delta$.   The CCM model state used is a helical state obtained from the $x$-aligned N\'{e}el state by applying an infinitesimal twist angle per unit length to it so that all of the spins lie in the $xy$ plane (and see text for details).  We show CCM LSUB$m$ results for $m=4,6,8,10$, together with the extrapolated LSUB$\infty(1)$ results based on this data set.  We also show the corresponding classical result from Eq.\ (\ref{rho_classical_eq}) with $s=\frac{1}{2}$.}
\end{figure}

Our CCM results for the spin stiffness coefficient $\rho_{s}$ are
shown in Fig.\ \ref{XXZ_fig4} in LSUB$m$ approximation
levels $m=4,6,8,10$, together with the corresponding LSUB$\infty(1)$
extrapolation based on this data set, over the range of values
$-1 \leq \Delta \leq 1$ of the anisotropy parameter.  
Again, as
expected, the results are exact in the $\Delta = -1$ limit.
Figure \ref{XXZ_fig4} shows the extremely rapid convergence of
the LSUB$m$ sequence of values for $\rho_{s}$ in the range
$-1 \leq \Delta \lesssim 0$, followed by a slower convergence in the
range $0 \lesssim \Delta \leq 1$.  The effect of quantum fluctuations
is again greatest in the vicinity of the isotropic Heisenberg point
($\Delta = 1$), where the difference from the classical result is
largest.

\begin{figure}[!t]
\center
\includegraphics[width=0.9\textwidth]{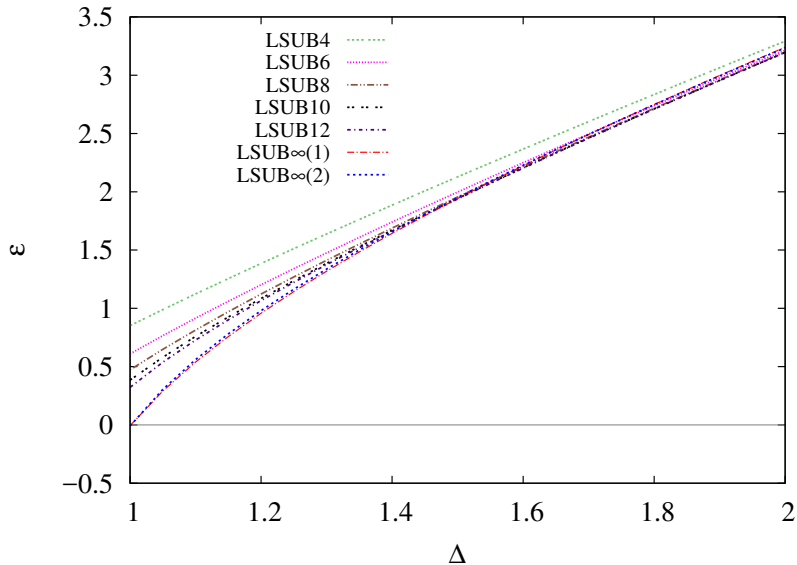}
\caption{\label{XXZ_fig5} The spin gap $\varepsilon$ for the spin-$\frac{1}{2}$ $XXZ$ antiferromagnet on the square lattice as a function of the anisotropy parameter $\Delta$.   The CCM ground-state model states used are N\'{e}el states aligned respectively in the $x$ direction for $-1 \leq \Delta \leq 1$ and in the $z$ direction for $\Delta \geq 1$.  We show CCM LSUB$m$ results for $m=4,6,8,10,12$, together with extrapolated LSUB$\infty(1)$ and LSUB$\infty(2)$ results based on the data sets $m=\{4,6,8,10\}$ and $m=\{4,6,8,10,12\}$, respectively.}
\end{figure}

Finally, in Fig.\ \ref{XXZ_fig5} we show our CCM results for the spin
gap $\varepsilon$ for a range of values $\Delta > 1$, where the system
is expected to be gapped.
Theoretically, we expect that $\varepsilon \rightarrow 0$ as the
isotropic Heisenberg limit $\Delta \rightarrow 1$ is approached and
the excitations become gapless Goldstone modes.  These modes then
persist for all values of the anisotropy parameter in the range
$-1 < \Delta \leq 1$, in which $\varepsilon$ remains zero.  From Fig.\
\ref{XXZ_fig5} we see that both LSUB$\infty$ extrapolations, based on
the two LSUB$m$ data sets $m=\{4,6,8,10\}$ and $m=\{4,6,8,10,12\}$,
give values of $\varepsilon$ at $\Delta = 1$ which are zero within
small numerical errors associated solely with the extrapolations.  The
actual LSUB$\infty$ extrapolated values at $\Delta = 1$ are
$\varepsilon = -0.0058$ using the LSUB$m$ data set $m=\{4,6,8,10\}$
and $\varepsilon = -0.0086$ using the LSUB$m$ data set
$m=\{4,6,8,10,12\}$.  One also observes from Fig.\ \ref{XXZ_fig5} that
the LSUB$m$ sequence of values $\varepsilon(m)$ for $\varepsilon$
converges appreciably more rapidly as $m \rightarrow \infty$ for
larger values of $\Delta$, and hence one expects that the associated
extrapolated values will be even more accurate than those obtained at
the $\Delta = 1$ limit.  Figure \ref{XXZ_fig5} shows that in the Ising limit, $\Delta \to \infty$, $\varepsilon$ becomes proportional to $\Delta$, exactly as expected classically.

\begin{figure}[!t]
\center
\includegraphics[width=0.8\textwidth]{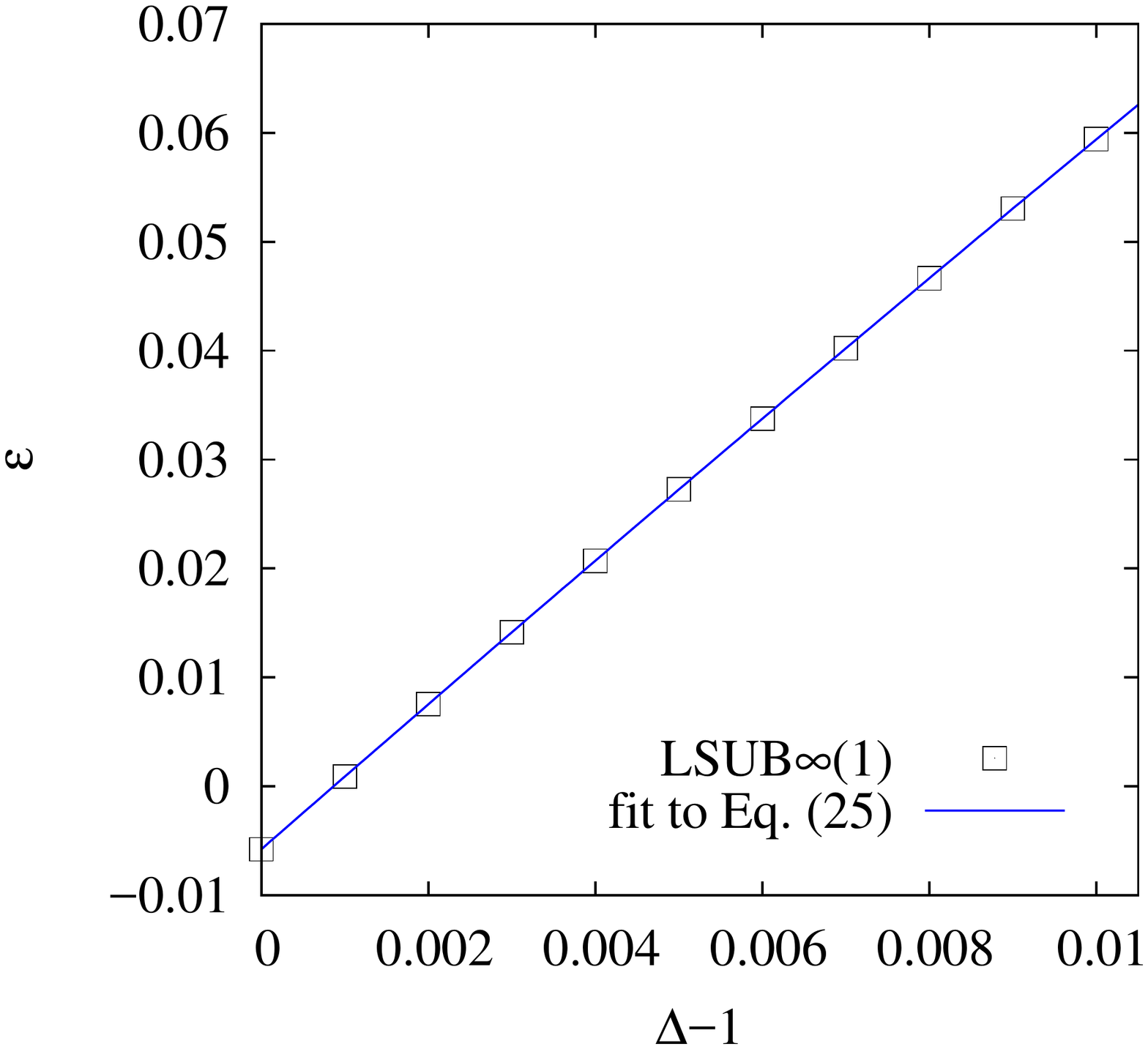}
\caption{\label{fit-GAP-near-heisenberg} The spin gap $\varepsilon$ for the spin-$\frac{1}{2}$ $XXZ$ antiferromagnet with anisotropy parameter $\Delta$ on the square lattice, plotted as a function of $(\Delta-1)$ in the vicinity of the Heisenberg point ($\Delta=1$).  The extrapolated LSUB$\infty(1)$ data points obtained from our CCM LSUB$m$ results based on the $z$-aligned N\'{e}el state as ground-state model state with $m=4,6,8,10$ are shown by open squares ($\Box$), and the solid line is the best fit to them of the form of Eq.\ (\ref{gap_fit}).}
\end{figure}

Once again, SWT predicts (and see, e.g., Ref.\ \cite{series}) however that
$\varepsilon$ vanishes near $\Delta = 1$ as
\begin{equation}
\frac{\varepsilon^{{\rm SWT}}}{\Delta} = \eta_{1}(1-\Delta^{-2})^{1/2} + \eta_{2}(1-\Delta^{-2}) + \eta_{3}(1-\Delta^{-2})^{3/2} + \cdots ~~ .  \label{spinGap_SWT}
\end{equation}
This behaviour, just as before for the ground-state parameters,
appears not to be borne out by our results shown in Fig.\
\ref{XXZ_fig5} for the spin-$\frac{1}{2}$ model.  To investigate
further we show in Fig.\ \ref{fit-GAP-near-heisenberg} our
extrapolated LSUB$\infty(1)$ results for the spin gap
$\varepsilon$ in the narrow range $1 \leq \Delta \leq 1.01$ based on
our LSUB$m$ results with $m=\{4,6,8,10\}$.  The leading (critical)
exponent $\nu$ is again obtained by fitting the extrapolated
LSUB$\infty(1)$ data points to the unbiased form
$\varepsilon=\epsilon_{0}+\epsilon_{1}(\Delta-1)^{\nu}$, where each of
the parameters $\epsilon_{0}$, $\epsilon_{1}$ and $\nu$ is fitted.
The best fit to the data points shown in Fig.\
\ref{fit-GAP-near-heisenberg} is obtained with
$\epsilon_{0}=-0.00584 \pm 0.00005$, $\epsilon_{1}=6.02 \pm 0.05$ and
$\nu=0.982 \pm 0.002$.  Once again, just as for the previous fits for
the staggered magnetization $M$ and the zero-field, uniform transverse
magnetic susceptibility $\chi$, the leading exponent $\nu$ takes a
fitted value $\nu$ very close to unity.  Hence, we now attempt a fit of
the form,
\begin{equation}
\varepsilon=\gamma_{0}+\gamma_{1}(\Delta-1)+\gamma_{2}(\Delta-1)^{2}  ~~ ,
\label{gap_fit}
\end{equation}
with $\gamma_{0}$ fixed at the value $-0.005774$ appropriate to the
LSUB$\infty(1)$ value for $\Delta=1$, obtained as described above
using the LSUB$m$ date set with $m=\{4,6,8,10\}$.  The best fit, shown
in Fig.\ \ref{fit-GAP-near-heisenberg} by the solid line, is obtained
with the values $\gamma_{1}=6.6941 \pm 0.0008$ and
$\gamma_{2}=-17.6 \pm 0.1$.

In Table \ref{tab1} we present our best CCM extrapolated
(LSUB$\infty$) results for each of the ground-state parameters
$e_{0}$, $M$, $\chi$ and $\rho_{0}$, together with the spin gap
$\varepsilon$, for various values of the anisotropy parameter
$\Delta$, in both the easy-axis $(\Delta > 1)$ and easy-plane
$(-1 < \Delta < 1)$ regimes, as well as at the isotropic
Heisenberg point $(\Delta=1)$.
\begin{table*}
\caption{Extrapolated CCM results for the ground-state energy per site $e_0$, the sublattice magnetization $M$ and the spin gap $\varepsilon$ are obtained for various values of the anisotropy parameter $\Delta$ by using the LSUB$m$ data set $m=\{4,6,8,10,12\}$. Extrapolated results for the zero-field transverse
susceptibility $\chi$, and the spin stiffness $\rho_s$ are obtained by using the LSUB$m$ data set $m=\{4,6,8,10\}$. The spin-wave velocity $c$ for the isotropic and easy-plane systems ($-1 \leq \Delta \leq 1$) can also be obtained by using the standard hydrodynamic relation $c^²=\sqrt{\rho_s/\chi}$.}
\vspace{-0.4cm}
\begin{center}
\setlength{\tabcolsep}{0.0pt} 
\renewcommand{\arraystretch}{1.4} 
{\footnotesize
\begin{tabular}{|c|c|c|c|c||c|c|c|c|c|c|c|c|}  \hline
$\Delta $  $\;$ & $\;$    $e_0$        $\;$ & $\;$  $M$          $\;$ & $\;$   $\chi$ $\;$ & $\;$    $\rho_s$       $\;$ & $\;$ $\Delta$ $\;$ & $\;$       $e_0$       $\;$ & $\;$    $M$      $\;$ & $\;$     $\chi$   $\;$ & $\;$ $\varepsilon$   $\;$ \\ \hline\hline
-1.00 $\;$ & $\;$       -0.5000      $\;$ & $\;$   0.5000    $\;$ & $\;$     $\infty$ $\;$ & $\;$   0.2500          $\;$ & $\;$  1.00    $\;$ & $\;$      -0.6696      $\;$ & $\;$   0.3093    $\;$ & $\;$     0.0691   $\;$ & $\;$  -0.0086    $\;$ \\ \hline    
-0.90 $\;$ & $\;$       -0.5010      $\;$ & $\;$   0.4924    $\;$ & $\;$     2.4154   $\;$ & $\;$   0.2581          $\;$ & $\;$  1.10    $\;$ & $\;$      -0.7028      $\;$ & $\;$   0.3766    $\;$ & $\;$     0.0783   $\;$ & $\;$   0.5601    $\;$ \\ \hline       
-0.80 $\;$ & $\;$       -0.5033      $\;$ & $\;$   0.4856    $\;$ & $\;$     1.1820   $\;$ & $\;$   0.2624          $\;$ & $\;$  1.15    $\;$ & $\;$      -0.7208      $\;$ & $\;$   0.3939    $\;$ & $\;$     0.0798   $\;$ & $\;$   0.7811    $\;$ \\ \hline       
-0.70 $\;$ & $\;$       -0.5066      $\;$ & $\;$   0.4792    $\;$ & $\;$     0.7738   $\;$ & $\;$   0.2654          $\;$ & $\;$  1.20    $\;$ & $\;$      -0.7394      $\;$ & $\;$   0.4067    $\;$ & $\;$     0.0805   $\;$ & $\;$   0.9805    $\;$ \\ \hline       
-0.60 $\;$ & $\;$       -0.5106      $\;$ & $\;$   0.4729    $\;$ & $\;$     0.5711   $\;$ & $\;$   0.2676          $\;$ & $\;$  1.25    $\;$ & $\;$      -0.7587      $\;$ & $\;$   0.4168    $\;$ & $\;$     0.0807   $\;$ & $\;$   1.1646    $\;$ \\ \hline       
-0.50 $\;$ & $\;$       -0.5154      $\;$ & $\;$   0.4667    $\;$ & $\;$     0.4502   $\;$ & $\;$   0.2692          $\;$ & $\;$  1.30    $\;$ & $\;$      -0.7784      $\;$ & $\;$   0.4249    $\;$ & $\;$     0.0805   $\;$ & $\;$   1.3371   $\;$ \\ \hline       
-0.40 $\;$ & $\;$       -0.5208      $\;$ & $\;$   0.4604    $\;$ & $\;$     0.3693   $\;$ & $\;$   0.2703          $\;$ & $\;$  1.35    $\;$ & $\;$      -0.7986      $\;$ & $\;$   0.4317    $\;$ & $\;$     0.0801   $\;$ & $\;$   1.5004   $\;$ \\ \hline 
-0.30 $\;$ & $\;$       -0.5269      $\;$ & $\;$   0.4542    $\;$ & $\;$     0.3123   $\;$ & $\;$   0.2708          $\;$ & $\;$  1.40    $\;$ & $\;$      -0.8191      $\;$ & $\;$   0.4374    $\;$ & $\;$     0.0796   $\;$ & $\;$   1.6563   $\;$ \\ \hline      
-0.20 $\;$ & $\;$       -0.5336      $\;$ & $\;$   0.4478    $\;$ & $\;$     0.2692   $\;$ & $\;$   0.2709          $\;$ & $\;$  1.50    $\;$ & $\;$      -0.8611      $\;$ & $\;$   0.4466    $\;$ & $\;$     0.0782   $\;$ & $\;$   1.9509   $\;$ \\ \hline 
-0.10 $\;$ & $\;$       -0.5409      $\;$ & $\;$   0.4413    $\;$ & $\;$     0.2358   $\;$ & $\;$   0.2706          $\;$ & $\;$  1.60    $\;$ & $\;$      -0.9041      $\;$ & $\;$   0.4537    $\;$ & $\;$     0.0767   $\;$ & $\;$   2.2279    $\;$ \\ \hline      
 0.00 $\;$ & $\;$       -0.5489      $\;$ & $\;$   0.4346    $\;$ & $\;$     0.2090   $\;$ & $\;$   0.2698          $\;$ & $\;$  1.70    $\;$ & $\;$      -0.9480      $\;$ & $\;$   0.4594    $\;$ & $\;$     0.0750   $\;$ & $\;$   2.4921   $\;$ \\ \hline      
 0.10 $\;$ & $\;$       -0.5575      $\;$ & $\;$   0.4276    $\;$ & $\;$     0.1870   $\;$ & $\;$   0.2685          $\;$ & $\;$  1.80    $\;$ & $\;$      -0.9925      $\;$ & $\;$   0.4641    $\;$ & $\;$     0.0733   $\;$ & $\;$   2.7465    $\;$ \\ \hline       
 0.20 $\;$ & $\;$       -0.5667      $\;$ & $\;$   0.4204    $\;$ & $\;$     0.1687   $\;$ & $\;$   0.2666          $\;$ & $\;$  1.90    $\;$ & $\;$      -1.0377      $\;$ & $\;$   0.4680    $\;$ & $\;$     0.0717   $\;$ & $\;$   2.9934   $\;$ \\ \hline      
 0.30 $\;$ & $\;$       -0.5766      $\;$ & $\;$   0.4128    $\;$ & $\;$     0.1531   $\;$ & $\;$   0.2640          $\;$ & $\;$  2.00    $\;$ & $\;$      -1.0833      $\;$ & $\;$   0.4712    $\;$ & $\;$     0.0700   $\;$ & $\;$   3.2344   $\;$ \\ \hline         
 0.40 $\;$ & $\;$       -0.5872      $\;$ & $\;$   0.4047    $\;$ & $\;$     0.1395   $\;$ & $\;$   0.2606          $\;$ & $\;$  2.50    $\;$ & $\;$      -1.3166      $\;$ & $\;$   0.4818    $\;$ & $\;$     0.0623   $\;$ & $\;$   4.3828   $\;$ \\ \hline        
 0.50 $\;$ & $\;$       -0.5985      $\;$ & $\;$   0.3960    $\;$ & $\;$     0.1276   $\;$ & $\;$   0.2562          $\;$ & $\;$  3.00    $\;$ & $\;$      -1.5555      $\;$ & $\;$   0.4875    $\;$ & $\;$     0.0559   $\;$ & $\;$   5.4790    $\;$ \\ \hline        
 0.60 $\;$ & $\;$       -0.6106      $\;$ & $\;$   0.3864    $\;$ & $\;$     0.1167   $\;$ & $\;$   0.2505          $\;$ & $\;$  3.50    $\;$ & $\;$      -1.7976      $\;$ & $\;$   0.4908    $\;$ & $\;$     0.0505   $\;$ & $\;$   6.5481    $\;$ \\ \hline     
 0.65 $\;$ & $\;$       -0.6169      $\;$ & $\;$   0.3811    $\;$ & $\;$     0.1115   $\;$ & $\;$   0.2469          $\;$ & $\;$  4.00    $\;$ & $\;$      -2.0417      $\;$ & $\;$   0.4930    $\;$ & $\;$     0.0460   $\;$ & $\;$   7.6008    $\;$ \\ \hline  
 0.70 $\;$ & $\;$       -0.6235      $\;$ & $\;$   0.3754    $\;$ & $\;$     0.1065   $\;$ & $\;$   0.2428          $\;$ & $\;$  4.50    $\;$ & $\;$      -2.2870      $\;$ & $\;$   0.4945    $\;$ & $\;$     0.0422   $\;$ & $\;$   8.6426    $\;$ \\ \hline       
 0.75 $\;$ & $\;$       -0.6304      $\;$ & $\;$   0.3692    $\;$ & $\;$     0.1016   $\;$ & $\;$   0.2380          $\;$ & $\;$  5.00    $\;$ & $\;$      -2.5333      $\;$ & $\;$   0.4955    $\;$ & $\;$     0.0390   $\;$ & $\;$   9.6764   $\;$ \\ \hline       
 0.80 $\;$ & $\;$       -0.6375      $\;$ & $\;$   0.3621    $\;$ & $\;$     0.0964   $\;$ & $\;$   0.2322          $\;$ & $\;$  6.00    $\;$ & $\;$      -3.0278      $\;$ & $\;$   0.4969    $\;$ & $\;$     0.0338   $\;$ & $\;$  11.7281     $\;$ \\ \hline       
 0.85 $\;$ & $\;$       -0.6449      $\;$ & $\;$   0.3540    $\;$ & $\;$     0.0911   $\;$ & $\;$   0.2249          $\;$ & $\;$  7.00    $\;$ & $\;$      -3.5238      $\;$ & $\;$   0.4977    $\;$ & $\;$     0.0298   $\;$ & $\;$  13.7657     $\;$ \\ \hline      
 0.90 $\;$ & $\;$       -0.6527      $\;$ & $\;$   0.3440    $\;$ & $\;$     0.0852   $\;$ & $\;$   0.2155          $\;$ & $\;$  8.00    $\;$ & $\;$      -4.0208      $\;$ & $\;$   0.4983    $\;$ & $\;$     0.0267   $\;$ & $\;$  15.7943     $\;$ \\ \hline      
 0.95 $\;$ & $\;$       -0.6609      $\;$ & $\;$   0.3306    $\;$ & $\;$     0.0784   $\;$ & $\;$   0.2022          $\;$ & $\;$  9.00    $\;$ & $\;$      -4.5185      $\;$ & $\;$   0.4986    $\;$ & $\;$     0.0241   $\;$ & $\;$  17.8167     $\;$ \\ \hline      
 1.00 $\;$ & $\;$       -0.6696      $\;$ & $\;$   0.3093    $\;$ & $\;$     0.0691   $\;$ & $\;$   0.1807          $\;$ & $\;$ 10.00    $\;$ & $\;$      -5.0167      $\;$ & $\;$   0.4989    $\;$ & $\;$     0.0220   $\;$ & $\;$  19.8347     $\;$ \\ \hline     
                                                                                                                           
\end{tabular}                                                                                              
}                                                                                              
\end{center}                                                                                
\label{tab1}                                                                                                
\end{table*}    
This tabulation should hence allow a direct comparison of our results
both to those obtained in appropriate experiments on systems to which
the model is applicable and in other theoretical approaches or
simulations using alternative techniques.

Before proceeding it is useful to compare our results to those
obtained by other approximate techniques for the two special cases
$\Delta=1$ and $\Delta=0$ of the anisotropy parameters.  Several
different techniques have been applied to study the spin-$\frac{1}{2}$
$XXZ$ model on the square lattice for $\Delta \geq 1$ (see, e.g.,
Refs.\
\cite{qmc5,series,thirdorderswt,Hamer:1994_SqLatt,weihong1995,witte1997,Takahasi1997}).
Both ED and QMC methods have also been applied to it in the range
$-1 \leq \Delta \leq 1$ (see, e.g., Ref.\ \cite{qmc5}).  Furthermore,
other techniques have also been applied for the specific case
($\Delta = 1$) of the isotropic Heisenberg model.  Our result for the
ground-state energy at $\Delta = 1$ is $e_{0} = -0.66964$.  This may
be compared firstly, for example, with corresponding results from
three different QMC simulations.  Thus, a zero-temperature ($T = 0$)
Green's function Monte Carlo (GFMC) calculation \cite{qmc3} directly
for the ground state gave $e_{0} = -0.66934(3)$, while another finite-temperature ($T \neq 0$) calculation using the stochastic series
expansion QMC (SSE-QMC) method \cite{qmc4} gave
$e_{0} = -0.699437(5)$.  Both of these calculations were performed on
$L \times L$ square lattices with $L \leq 16$, and the results
extrapolated to the thermodynamic limit ($L \rightarrow \infty$).  Two
other $T \neq 0$ QMC simulations of the model, based on a continuous
Euclidean time version of a loop cluster algorithm for evaluating path
integrals (PIMC) \cite{Beard:1996_SqLatt,Kim:1998_SqLatt_QMC},
extracted the low-energy parameters by fitting the $T \neq 0$ data to
finite-temperature scaling forms derived from chiral perturbation
theory \cite{Hasenfratz:1993_magnon-chiral-PT}.  Using very
large-scale simulations on $L \times L$ lattices with $L \leq 1000$,
for example, Kim and Troyer \cite{Kim:1998_SqLatt_QMC} found
$e_{0} = -0.66953(4)$.  The spin-$\frac{1}{2}$ isotropic Heisenberg
model on the square lattice has also been studied via extrapolations
to the $N \rightarrow \infty$ limit on ED calculations of clusters of
sizes $N \leq 40$ \cite{ED40}, which gave a ground-state energy
$e_{0} = -0.6701$; and by extrapolations to the $\Delta = 1$ limit
using a linked-cluster SE method around the Ising
($\Delta \rightarrow \infty$) limit \cite{series}, which gave
$e_{0} = -0.6693(1)$.  It is clear that our CCM result for the
ground-state energy at $\Delta = 1$ is in complete agreement with
these other accurate results.  For comparison purposes the
corresponding result at $\Delta = 1$ from SWT \cite{thirdorderswt} up
to third order in powers of $1/s$ about the classical
($s \rightarrow \infty$) limit from Eq.\ (\ref{E_classical}) is given
by
\begin{equation}
e_{0} = -2s^{2} - 0.315895s - 0.012474 + 0.000216(6)s^{-1} + {\cal O}(s^{-2}) ~~ .  \label{SWT-3rdOrd_classical-infty_E}
\end{equation}
For our present $s=\frac{1}{2}$ model Eq.\
(\ref{SWT-3rdOrd_classical-infty_E}) yields the respective
approximations at first, second and third orders in SWT,
$e_{0} = -0.65795$ (SWT1), $e_{0} = -0.67042$ (SWT2), and
$e_{0} = -0.66999$ (SWT3).

For the $\Delta = 1$ case our CCM result for the order parameter is
$M = 0.3093$.  Once again, this may be compared with a $T \neq 0$ SSE-QMC
result \cite{qmc4} of $M=0.3070(3)$ and the extrapolated result from
a $T \neq 0$ PIMC calculation \cite{Beard:1996_SqLatt} of $M = 0.3083(2)$.  A further study of
the present $XXZ$ model using a combination of ED and QMC results
\cite{qmc5} gave $M=0.3050(5)$ for the case $\Delta = 1$, while a
direct extrapolation to the $N \rightarrow \infty$ limit on ED
calculations with clusters of sizes $N \leq 40$ \cite{ED40} gave the
results $M=0.3105$.  Lastly, the corresponding result from a
linked-cluster SE method around the Ising limit
($\Delta \rightarrow \infty$) \cite{series}, suitably extrapolated to
the $\Delta = 1$ limit, gave the value $M=0.307(1)$.  Once again, we
see that our CCM result for the ground-state order parameter $M$ at
$\Delta = 1$ agrees well with these other accurate
results.  Again, for purposes of comparison, the result from SWT up to
third order \cite{thirdorderswt,Igarashi:1992_SqLatt} for the $\Delta = 1$ case about the
classical ($s \rightarrow \infty$) result of $M = s$ is given by
\begin{equation}
M = s - 0.1966019 + 0.00087(1)s^{-2} + {\cal O}(s^{-3}) ~~ . \label{SWT-3rdOrd_classical-infty_M}
\end{equation}
For our present $s=\frac{1}{2}$ model Eq.\
(\ref{SWT-3rdOrd_classical-infty_M}) yields the respective
approximations at first, second and third orders in SWT, $M=0.3034$
(SWT1), $M=0.3034$ (SWT2), and $M=0.3069$ (SWT3).

Our results for the spin stiffness and zero-field, uniform transverse magnetic
susceptibility at $\Delta = 1$ are $\rho_{s} = 0.1807$ and
$\chi = 0.0691$, respectively.  These may firstly be compared with the
results of various ED and QMC calculations.  For example, a study
using a combination of ED and QMC results \cite{qmc5} gave values
$\rho_{s} = 0.180(2)$ and $\chi = 0.0755(15)$ in the thermodynamic
limit, while a direct extrapolation to the $N \rightarrow \infty$
limit on ED calculations of clusters of sizes $N \leq 40$ \cite{ED40}
gave the two different values $\rho_{s}=0.1246$ extracted from the
finite-size scaling relation for the order parameter, and
$\rho_{s}=0.1115$ extracted from the value $\chi=0.0674$ obtained for the zero-field, uniform transverse magnetic susceptibility and the corresponding value $c=1.287$ obtained for the spin-wave velocity
$c$, together with the hydrodynamic relation (see, e.g., Refs.\
\cite{Hasenfratz:1993_magnon-chiral-PT,Halperin:1969_SWT,Chakravarty:1989_magnon-chiral-PT,Cherryshev:2009_SWT}),
\begin{equation}
\rho_{s} = \chi c^{2} ~~ ,  \label{sStiff_hydrodynmamic_SWTc}
\end{equation}
which is valid both for Heisenberg and general easy-plane
antiferromagnets.  Both parameters were also calculated directly in a
$T \neq 0$ SSE-QMC simulation of the isotropic ($\Delta=1$) model
\cite{qmc4}, which gave values $\rho_{s}=0.175(2)$ and
$\chi=0.0625(9)$.  By contrast, a $T=0$ GFMC simulation of the
isotropic ($\Delta=1$) model \cite{qmc3} calculated $\chi$ and $c$
directly.  Use of Eq.\ (\ref{sStiff_hydrodynmamic_SWTc}) enables us to
quote the corresponding GFMC results $\rho_{s}=0.162(10)$ and
$\chi=0.0669(7)$.  Two $T \neq 0$ PIMC simulations of the isotropic
system may also be quoted.  The first \cite{Beard:1996_SqLatt} finds
$\rho_{s}=0.185(2)$ and quotes a value $c=1.68(1)$, from which we find
$\chi=0.0655(15)$.  By contrast, a second very large-scale PIMC
simulation \cite{Kim:1998_SqLatt_QMC} calculates both $\rho_{s}$ and
$\chi$ directly, and quotes the values $\rho_{s}=0.178(2)$ and
$\rho_{s}=0.185(1)$ from two different fits to the data, and
$\chi=0.06549(2)$.  Lastly, the corresponding values obtained directly
from a linked-cluster SE method around the Ising limit
($\Delta \rightarrow \infty$), suitably extrapolated to the
$\Delta = 1$ limit, are $\rho_{s}=0.182(5)$ \cite{Hamer:1994_SqLatt}
and $\chi = 0.0659(10)$ \cite{series}.  We see once more that our CCM
results for both $\rho$ and $\chi$ at $\Delta = 1$ are in very good
agreement with other purportedly accurate results.

Again, for comparison, we also cite corresponding results from SWT for the $\Delta = 1$ case.  For the spin stiffness results are known \cite{Hamer:1994_SqLatt} up to third order in powers of $1/s$ about the classical ($s \rightarrow \infty$) limit from Eq.\ (\ref{rho_classical_eq}),
\begin{equation}
\rho_{s} = s^{2}-0.117629s-0.010208-0.00316(2)s^{-1}+{\cal O}(s^{-2}) ~~ .  \label{rho_class-limit_results}
\end{equation}
Corresponding results for $\chi$ are known at $\Delta =1$ \cite{Igarashi:1992_SqLatt,Hamer:1994_SqLatt} up to second order in powers of $1/s$ about the classical limit from Eq.\ (\ref{chi_classical_eq}),
\begin{equation}
\chi = 0.125-0.034447s^{-1}+0.002040s^{-2}+{\cal O}(s^{-3}) ~~ .   \label{chi_class-limit_results}
\end{equation}
Note that the term proportional to $s^{-2}$ in the SWT expansion for
$\chi$ in Ref.\ \cite{thirdorderswt} was later corrected in Ref.\
\cite{Hamer:1994_SqLatt} to that shown in Eq.\
(\ref{chi_class-limit_results}).  Equations
(\ref{rho_class-limit_results}) and (\ref{chi_class-limit_results})
yield for our present $s=\frac{1}{2}$ model the respective
approximations at first, second and third orders in SWT,
$\rho_{s}=0.1912$ (SWT1), $\rho_{s}=0.1810$ (SWT2), $\rho_{s}=0.1747$
(SWT3), and $\chi=0.0561$ (SWT1), and $\chi = 0.0643$ (SWT2).

For the corresponding case $\Delta=0$ of the anisotropy parameter,
which equates to the spin-$\frac{1}{2}$ isotropic $XY$ ($\equiv XX$)
model, our CCM results are $e_{0}=-0.54888$, $M=0.4346$,
$\rho_{s}=0.2698$ and $\chi=0.2090$.  These may be compared with
results from a study using a combination of ED and QMC results
\cite{qmc5}, which gave $e_{0}=-0.54882(3)$, $M=0.4377(5)$,
$\rho_{s}=0.2695(2)$, and $\chi=0.211(1)$; and from a
finite-temperature ($T \neq 0$) SSE-QMC simulation \cite{xymc}, which
gave $e_{0}=-0.548824(2)$, $M=0.437(2)$, $\rho_{s}=0.2696(2)$ and
$\chi=0.2096(2)$.  Our results are thus again seen to be in very good
agreement with these other accurate results for the $\Delta=0$ case.

\begin{figure*}
\begin{center}
\mbox{
\hspace{-1.5cm}
\subfigure[]{\scalebox{0.72}{\includegraphics{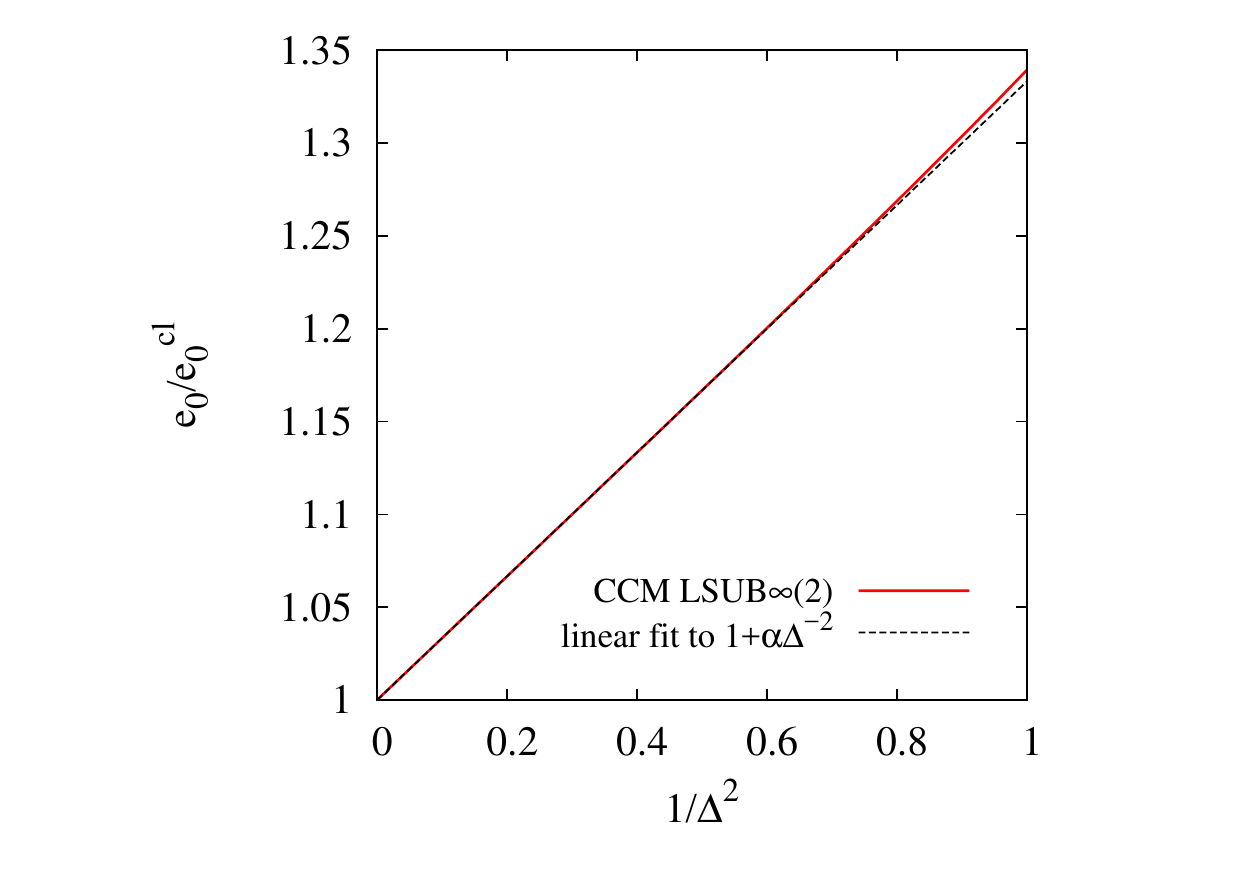}}}
\hspace{-2.0cm}
\subfigure[]{\scalebox{0.72}{\includegraphics{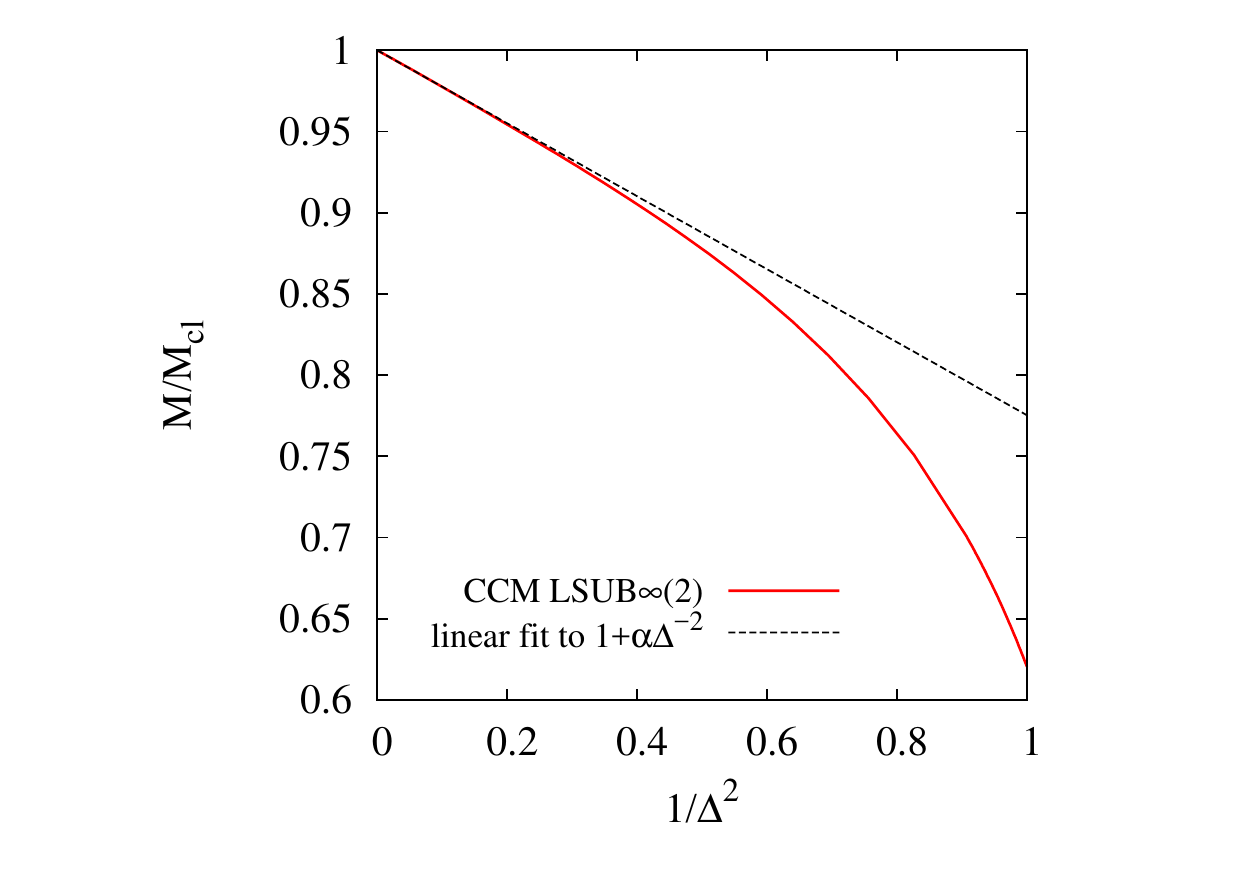}}}
}
\mbox{
\hspace{-1.5cm}
\subfigure[]{\scalebox{0.72}{\includegraphics{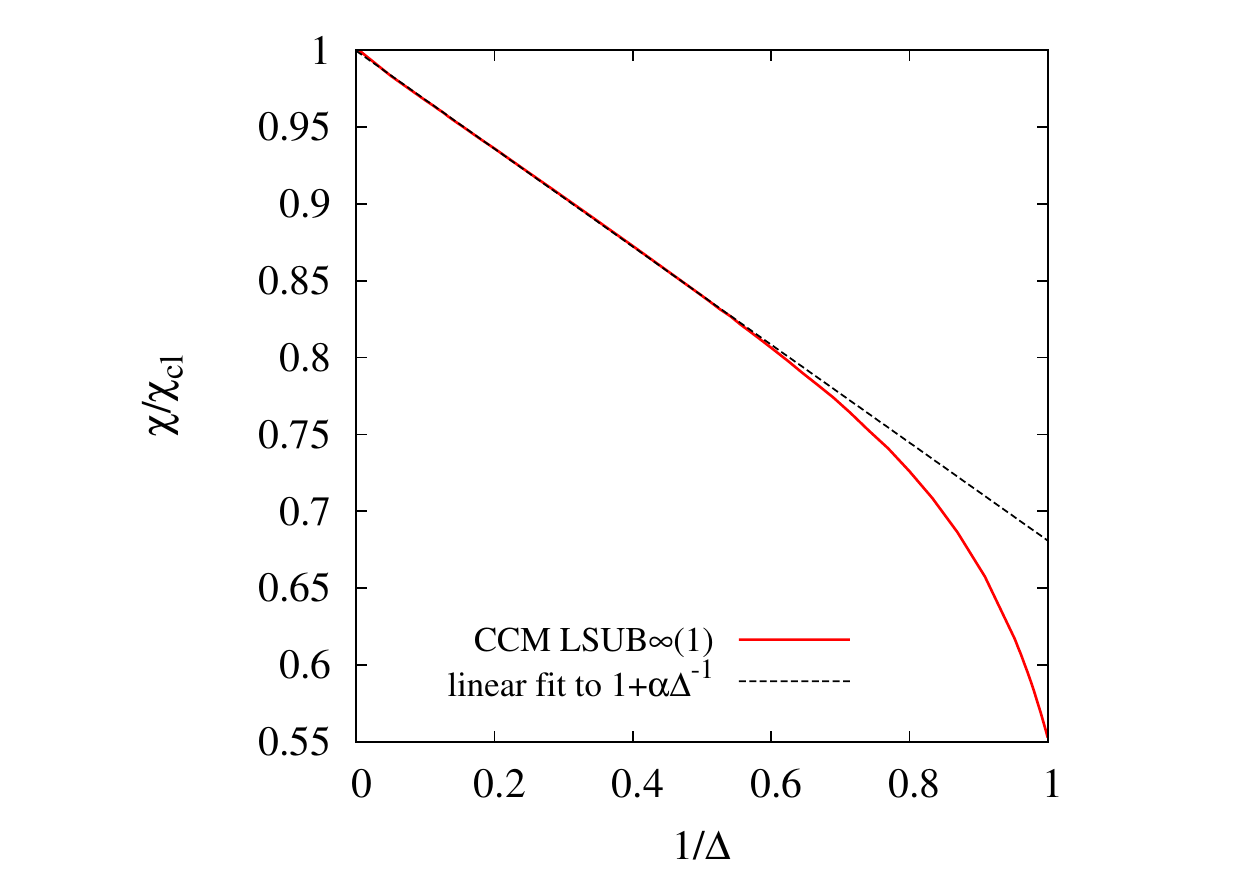}}}
\hspace{-2.0cm}
\subfigure[]{\scalebox{0.72}{\includegraphics{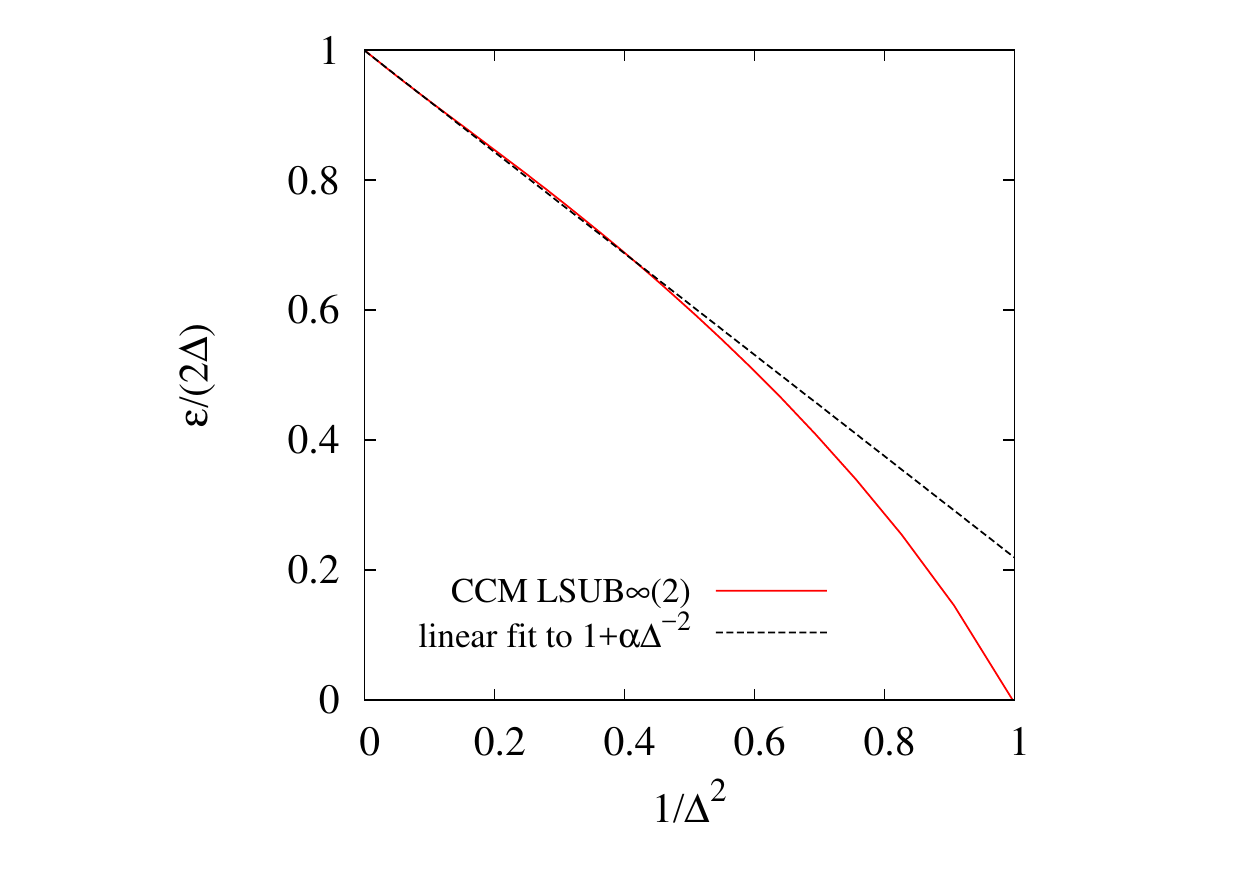}}}
}
\caption{Extrapolated CCM results for the (a) scaled ground-state energy per site $e_{0}/e^{{\rm cl}}_{0}$, (b) scaled sublattice magnetization
$M/M_{{\rm cl}}$, (c) scaled susceptibility $\chi/\chi_{{\rm cl}}$, and (d) scaled spin gap $\varepsilon/(2\Delta)$, plotted as functions of either $1/\Delta^{2}$ or $1/\Delta$, as shown, where $\Delta$ is the anisotropy parameter, in the region $\Delta \geq 1$.  The LSUB$\infty(1)$ results for $\chi/\chi_{{\rm cl}}$ are based on LSUB$m$ data with $m=\{4,6,8,10\}$, while the LSUB$\infty(2)$ results for the remaining quantities are based on LSUB$m$ data with $m=\{4,6,8,10,12\}$.   In each case we also show least-squares linear fits of the form $1 + \alpha\Delta^{-n}$, where $n=2$ (for $e_{0}/e_{0}^{{\rm cl}}$, $M/M_{{\rm cl}}$ and $\varepsilon/(2\Delta)$) and $n=1$ for $\chi/\chi_{{\rm cl}}$, to the CCM data points with $\Delta \geq 3$.}
\label{fig6behavior}
\end{center}
\end{figure*}

Finally, in Fig.\ \ref{fig6behavior} we present our extrapolated CCM results for the ground-state quantities $e_{0}$, $M$ and $\chi$ (in each case as a ratio with respect to their classical counterparts), as well as for the ratio $\varepsilon/(2\Delta)$, in the region $\Delta \geq 1$.
For reasons we describe below $e_{0}/e^{{\rm cl}}_{0}$,
$M/M_{{\rm cl}}$ and $\varepsilon/(2\Delta)$ are plotted as functions
of $1/\Delta^{2}$, while $\chi/\chi_{{\rm cl}}$ is plotted against
$1/\Delta$.  As expected, we observe that each of the four scaled
parameters approaches the value 1 in the Ising limit
($\Delta \rightarrow \infty$) where the CCM becomes exact.  It is interesting to compare our
results with those obtained from perturbation theory (PT) expansions
in powers of $1/\Delta$ around the Ising limit.  For the ground-state
energy the PT expansion \cite{series,Singh:1989} is
\begin{equation}
\frac{e_{0}}{e^{{\rm cl}}_{0}} = 1+\frac{1}{3}\frac{1}{\Delta^{2}} -\frac{1}{540}\frac{1}{\Delta^{4}}+{\cal O}(\frac{1}{\Delta^{6}}) ~~ ,  \label{GS_E_PT}
\end{equation}
while for the ground-state order parameter the corresponding PT expansion \cite{series,Singh:1989,huse88} is
\begin{equation}
\frac{M}{M_{{\rm cl}}} = 1-\frac{2}{9}\frac{1}{\Delta^{2}}-\frac{8}{225}\frac{1}{\Delta^{4}}+{\cal O}(\frac{1}{\Delta^{6}}) ~~ . \label{GS_M_PT}
\end{equation}
In Figs.\ \ref{fig6behavior}(a) and \ref{fig6behavior}(b) we also show
least-squares straight-line fits of the form $1+\alpha\Delta^{-2}$ to
the extrapolated CCM data points with $\Delta \geq 3$ for
$e_{0}/e^{{\rm cl}}_{0}$ and $M/M_{{\rm cl}}$.  For
$e_{0}/e^{{\rm cl}}_{0}$ we obtain a fit with $\alpha = 0.33320 \pm 0.00001$, which
may be compared with the exact value $\frac{1}{3}$ from Eq.\
(\ref{GS_E_PT}).  
The corresponding fitted value for
$M/M_{{\rm cl}}$ in Fig.\ \ref{fig6behavior}(b) is $\alpha=-0.2247 \pm 0.0002$,
which may be compared with the exact value $-\frac{2}{9}$ from Eq.\
(\ref{GS_M_PT}).  The corresponding PT series around the Ising limit for the zero-field,
uniform transverse susceptibility $\chi$ \cite{series,Singh:1989}
contains both odd and even powers of $\Delta^{-1}$,
\begin{equation}
\Delta\chi = \frac{1}{4}-\frac{1}{3}\frac{1}{\Delta}+\frac{17}{48}\frac{1}{\Delta^{2}}-\frac{41}{108}\frac{1}{\Delta^{3}}+{\cal O}(\frac{1}{\Delta^{4}}) ~~ ,  \label{Delta_chi_PT}
\end{equation}
unlike those for $e_{0}/\Delta$ and $M$, which contain only even powers of $\Delta^{-1}$.  Using Eq.\ (\ref{chi_classical_eq}), the corresponding expansion for $\chi/\chi_{{\rm cl}}$ is thus,
\begin{equation}
\frac{\chi}{\chi_{{\rm cl}}} = 1-\frac{1}{3}\frac{1}{\Delta}+\frac{1}{12}\frac{1}{\Delta^{2}}-\frac{11}{108}\frac{1}{\Delta^{3}}+{\cal O}(\frac{1}{\Delta^{4}}) ~~ .  \label{scaled_chi_PT}
\end{equation}
In Fig.\ \ref{fig6behavior}(c) we also show a least-squares
straight-line fit of the form $1+\alpha\Delta^{-1}$ to the
extrapolated CCM data points with $\Delta \geq 3$ for
$\chi/\chi_{{\rm cl}}$.  The obtained value is
$\alpha=-0.321 \pm 0.001$, which may be compared with the exact value
$-\frac{1}{3}$ from Eq.\ (\ref{scaled_chi_PT}).

Finally, the corresponding PT series around the Ising limit for the scaled spin
gap $\varepsilon/(2\Delta)$ is \cite{series},
\begin{equation}
\frac{\varepsilon}{2\Delta} = 1-\frac{5}{6}\frac{1}{\Delta^{2}}+\frac{137}{864}\frac{1}{\Delta^{4}}+{\cal O}(\frac{1}{\Delta^{6}}) ~~ ,  \label{spinGap_PT}
\end{equation}
which again contains only even powers of $\Delta^{-1}$.  The
least-squares fit, shown in Fig.\ \ref{fig6behavior}(d), of the form
$1+\alpha\Delta^{-2}$ to the CCM data points with $\Delta \geq 3$ for
$\varepsilon/(2\Delta)$ yields a value $\alpha=-0.794 \pm 0.003$, which may be compared with the exact
value $-\frac{5}{6}$ from Eq.\ (\ref{spinGap_PT}).  

\begin{figure}[!t] 
\center
\includegraphics[width=0.9\textwidth]{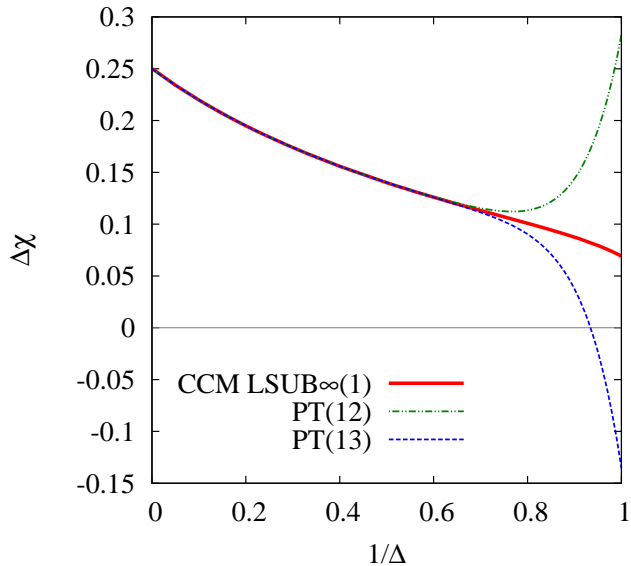}
\caption{\label{fig8XXZ} Results for the scaled zero-field, uniform transverse susceptibility $\Delta\chi$ as a function of the inverse anisotropy parameter $1/\Delta$, in the region $\Delta \geq 1$, from both our CCM LSUB$\infty(1)$ extrapolation using LSUB$m$ data with $m=\{4,6,8,10\}$ and the $n$th-order PT($n$) expansions about the Ising limit ($\Delta \rightarrow \infty$) with $n=12$ and $n=13$.} 
\end{figure}

It is interesting to note from Fig.\ \ref{fig6behavior} that even
lowest-order PT (i.e., the straight-line fits shown) gives rather
accurate results for each of the parameters shown for values of the
anisotropy parameter $\Delta \gtrsim 1.8$.  In each case in this range
the extrapolated CCM values and the straight-line fits are difficult
to distinguish by eye.  It is thus natural to ask how the inclusion of
additional terms in the PT expansions changes the accuracy of the
results for smaller values of $\Delta$ as we approach the isotropic
Heisenberg limit ($\Delta \rightarrow 1$).  Let us denote by PT($n$)
the corresponding $n$th-order PT series around the Ising
($\Delta \rightarrow \infty$) limit for the respective model parameter
under consideration, scaled to its classical (large-$\Delta$) value
(i.e., the series terminated at the term proportional to
$\Delta^{-n}$).  Such series expansions have been given, for example,
in Ref.\ \cite{series} for $e_{0}/e^{{\rm cl}}_{0}$ and
$M/M_{{\rm cl}}$ out to $n=14$, for $\chi/\chi_{{\rm cl}}$ out to
$n=13$, and for $\varepsilon/(2\Delta)$ out to $n=10$.  A similar
expansion for $\rho_{s}$ has been given, for example, in Ref.\
\cite{Hamer:1994_SqLatt} out to $n=10$.

Thus, in Fig.\ \ref{fig8XXZ} we take the specific example of the
zero-field transverse magnetic
susceptibility, where we compare results for the quantity $\chi\Delta$
from our own CCM extrapolation using LSUB$m$ data with
$m=\{4,6,8,10\}$ to those obtained from the PT$(13)$ expansion (i.e.,
as in Eq.\ (\ref{Delta_chi_PT}) but including 14 terms out to the term
proportional to $1/\Delta^{13}$).
The two curves are now essentially indistinguishable by eye for all
values $\Delta \gtrsim 1.4$ of the $XXZ$ model anisotropy parameter
$\Delta$.  Nevertheless, despite the extraordinarily close agreement
in this range, what is very interesting is how rapidly the two curves
diverge from one another as $\Delta$ is reduced further.  Whereas the CCM
results remain smooth even as $\Delta \rightarrow 1$, the PT$(13)$
results become wholly unphysical (i.e., $\chi < 0$) in this limit for
all values $1 \leq \Delta \lesssim 1.07$.  For comparison purposes we also show in Fig.\ \ref{fig8XXZ} the PT results at the PT(12) level.  We clearly observe that the PT series becomes ill-behaved as we approach the critical point at $\Delta=1$.  We comment further on these findings in
Sec.\ \ref{summ_sec} below.

\section{Summary and discussion}
\label{summ_sec}
The spin-half square-lattice {\it XXZ} antiferromagnet is a
fundamental and prototypical model of quantum magnetism, to which a
variety of quantum many-body theory techniques has previously been
applied.  In this paper we have applied the high-order CCM to the
model, using two reference (or model) states upon which to build the
multi-spin correlations in a fully consistent LSUB$m$ hierarchy.
Unlike most alternative techniques the CCM has the distinct advantage
that we work from the outset, at every level of LSUB$m$ approximation,
in the large-lattice ($N \rightarrow \infty$) thermodynamic limit.  We
have presented results for the ground-state energy, the sublattice
magnetization (i.e., the order parameter), the spin stiffness, the
zero-field, uniform transverse magnetic susceptibility, and the
triplet spin gap, for a large range of values of the $XXZ$ anisotropy
parameter $\Delta$.  The CCM results for each of these parameters
were found to converge rapidly with increasing values of the LSUB$m$
truncation parameter $m$, for all values of $\Delta$ (in the range
$-1\leq \Delta < \infty$ of interest), and we showed how simple
heuristic extrapolation schemes for $m \rightarrow \infty$ could be
used to estimate the formally exact LSUB$\infty$ values.

Our CCM LSUB$m$ results are exact in the two limits $\Delta = -1$
(where there is a first-order phase transition to a ferromagnetic
state) and $\Delta \rightarrow \infty$ (the Ising limit).  The most
interesting point in between these limits is at the isotropic
Heisenberg (or $XXX$) point, $\Delta = 1$, where the model possesses
SU(2) spin-rotational symmetry.  The ground state of the isotropic
model then undergoes spontaneous symmetry breaking via the Goldstone
mechanism, so that as the limit $\Delta \rightarrow 1$ is approached
from the Ising side ($\Delta > 1$) the system has long-range N\'{e}el
order in the $z$ direction with a predicted finite value of the
corresponding order parameter, $M \approx 0.309$.  We showed that in
the same limit $\Delta \rightarrow 1$, the spin gap vanishes
($\varepsilon \rightarrow 0$) within very small numerical errors,
corresponding to the emergence of the massless Goldstone boson
excitation modes.  Away from the isotropic limit, when
$\Delta \neq 1$, the SU(2) spin-rotational symmetry is broken into a
product of a Z(2) symmetry in the $z$ direction and a U(1) symmetry in
the $xy$ plane.

Precisely at the isotropic Heisenberg point ($\Delta=1$) all of the
parameters calculated exhibit the greatest difference from their
classical counterparts, and hence we expect any errors in our (and
other) calculations to be greatest for this value of $\Delta$.
However, we have shown specifically at $\Delta = 1$ that our results
compare extremely well with those from a number of different QMC
simulations, as well as with the results of linked-cluster SE
techniques and high-order SWT.  As expected, our results are even
closer to those of QMC simulations at the isotropic $XY$ (or $XX$)
point, $\Delta = 0$.  All of these results demonstrate very clearly
the high accuracy of which the CCM is capable.

We have exploited this accuracy to examine the behaviour of the model
parameters in the vicinity of the isotropic Heisenberg point,
$\Delta=1$.  Whereas SWT indicates that the point $\Delta=1$ is
singular, with the physical parameters behaving there as power series
in $(1-\Delta^{-2})^{1/2}$ on the Ising side, as in Eqs.\
(\ref{E-eq_SWT}), (\ref{M-eq_SWT}), (\ref{chi-eq_SWT}) and
(\ref{spinGap_SWT}), our own analysis of the sublattice magnetization
$M$, for example, gave a different value of the leading exponent
[c.f., Eqs.\ (\ref{M-eq_SWT}) and (\ref{M_fit})].  Similar analyses of
our CCM results for both the zero-field, uniform transverse magnetic
susceptibility $\chi$ [c.f., Eqs.\ (\ref{chi-eq_SWT}) and
(\ref{chi_fit})] and the triplet spin gap $\varepsilon$ [c.f., Eqs.\
(\ref{spinGap_SWT}) and (\ref{gap_fit})] of the spin-$\frac{1}{2}$
square-lattice $XXZ$ antiferromagnet in the easy-axis regime near the
singular isotropic point $\Delta=1$ also show marked differences from
the square-root singularities predicted by SWT.  Our CCM results for
all three parameters $p=\{M,\chi,\varepsilon\}$ in this critical regime
show a consistently different form of criticality to that predicted by
SWT .  In each case, if we attempt a fit to our LSUB$\infty$ results
of the form $p \to p_{0} + p_{1}(\Delta-1)^{\nu}$ as
$\Delta \to 1^{+}$ in the critical regime, we find a value of $\nu$ very
close to 1 rather than the value $\frac{1}{2}$ from SWT.  With a value
$\nu=1$ it is then also possible that the associated critical
behaviour is more subtle than a simple leading power law (e.g.,
involving additional logarithmic or other non-algebraic terms).  Since
the behaviour of the model parameters near $\Delta=1$ predicted by SWT
presumably becomes exact in the $s \to \infty$ limit, the intriguing
possibility opens up that the leading critical exponent describing the
singular behaviour there depends on the spin quantum number $s$.  Any
further such analysis is beyond the bounds of the present paper,
however.

These results are particularly
interesting in the context that the PT($n$) perturbative power series
expansions about the Ising limit [and see, e.g., Eqs.\
(\ref{GS_E_PT})--(\ref{scaled_chi_PT})] are very ill-behaved near
$\Delta=1$, as is to be expected, and as Fig.\ \ref{fig8XXZ} shows for the transverse
susceptibility $\chi$, for example.  In order to extrapolate these
PT$(n)$ series to the isotropic limit it is necessary to make some
appropriate analytic continuation, and the approximate methods to do
so lie at the heart of all linked-cluster SE approaches (and see, e.g.,
Refs.\
\cite{series,Hamer:1994_SqLatt,Singh:1989,weihong1995,witte1997}).
For example, in the present case, it is usual (and see, e.g., Refs.\
\cite{series,Hamer:1994_SqLatt,Singh:1989}) to first transform the
PT($n$) series in $\frac{1}{\Delta}$ to a new variable
$\delta \equiv 1-(1-\Delta^{-2})^{1/2}$, so that according to SWT the
series should then be analytic in $\delta$.  The $\delta$-series is
then extrapolated to the point $\delta=1$ by some suitable (e.g.,
Pad\'{e} or an integrated first-order inhomogeneous differential)
approximant.  Clearly, the extrapolated values so obtained do depend on
the assumptions about the singularity exponents, which are numerically
only very poorly determined by the series themselves.

It is worth pointing out that our CCM results based on the $z$-aligned
N\'{e}el model state at the LSUB$m$ level of approximation reproduce
exactly the large-$\Delta$ perturbative expansions at the same PT($m$)
order.  Whereas such PT($n$) expansions are generally calculated by
linked-cluster techniques (and see, e.g., Ref.\
\cite{He:1990_SqLatt}), the linked-cluster SE method that utilizes
them (and see, e.g., Refs.\
\cite{series,Hamer:1994_SqLatt,Singh:1989}) must then use appropriate
extrapolation methods to evaluate the series at the required parameter value (e.g.,
$\Delta = 1$ for the isotropic Heisenberg model).  Other similar
methods, such as the $t$-expansion method
\cite{Horn:1984_t-expansion}, the connected-moments expansion (CMX)
method \cite{Cioslowski:1987_CMX_PRL,Cioslowski:1987_CMX_PRA} and the
(plaquette expansion or) analytic Lanczos expansion (ALE) method
\cite{Hollenberg:1993_ALE,Witte:1994_ALE}, each of which has also been
applied to the present model \cite{weihong1995,witte1997}, also
require similar extrapolations to be performed.  Each of these methods
(viz., the linked-cluster SE, the $t$-expansion, the CMX and the ALE
methods) shares with the CCM, however, that they are all based on
linked-cluster theorems, such that thermodynamically extensive
variables, such as the ground-state energy, can be computed in terms
of connected diagrams.  The strength of the CCM is that it both works
directly in the large-lattice ($N \rightarrow \infty$) lattice from
the outset at all LSUB$m$ levels of approximation, and that it never
needs to extrapolate any intrinsically perturbative series.  Since it
is well known that any uncertainties in the knowledge of the global
analytic properties of such series are usually the biggest source of
poor convergence and associated errors, the CCM has a unique advantage
over these other methods in this regard.

In conclusion, we have provided results for this prototypical model of
quantum magnetism over a wide range of values of the anisotropy
parameter in both graphical and tabular formats, in order to
facilitate their quantitative comparison with those from other
approximate methods and from experiment.  We hope that the CCM results
presented here will thus provide a useful yardstick for both theorists
and experimentalists studying related magnetic materials.

\section*{Acknowledgements}
One of us (R.F.B.) gratefully acknowledges the Leverhulme
Trust (United Kingdom) for the award of an Emeritus Fellowship (EM-2015-007).  We also thank Oliver G\"{o}tze for valuable discussions.


\section*{References}








\end{document}